\documentclass[prl,aps,twocolumn,10pt,superscriptaddress,notitlepage,longbibliography]{revtex4-2}
\usepackage[dvipsnames]{xcolor}
\usepackage[english]{babel} 
\usepackage{graphicx}
\usepackage{float}
\usepackage{braket}
\usepackage{epstopdf}
\usepackage{mathtools}
\usepackage{amsmath}
\usepackage{amssymb}
\usepackage{bbm,bm}
\usepackage[caption=false]{subfig}
\usepackage{pythonhighlight}
\usepackage{float}
\usepackage{bbold}
\usepackage[T1]{fontenc}

\usepackage{hyperref}
 \hypersetup{
     colorlinks=true,
     linkcolor=blue,
     filecolor=blue,
     citecolor = magenta,      
     urlcolor=red,
     }

\usepackage[markup=underlined]{changes}
\makeatletter
\@namedef{Changes@AuthorColor}{magenta}
\colorlet{Changes@Color}{magenta}
\makeatother
\usepackage{hyperref}
 \hypersetup{
     colorlinks=true,
     linkcolor=blue,
     filecolor=blue,
     citecolor = magenta,      
     urlcolor=red,
     }

\usepackage{braket}

\usepackage{xargs}
\newcommandx{\greencom}[2][1=]
{\todo[inline, color=green!40,#1]{#2}}
\newcommandx{\bluecom}[2][1=]
{\todo[inline, color=blue!40,#1]{#2}}
\newcommandx{\bluemargin}[2][1=]
{\todo[color=blue!40,#1]{#2}}

\usepackage{letltxmacro}
\LetLtxMacro{\ORIGselectlanguage}{\selectlanguage}
\makeatletter
\DeclareRobustCommand{\selectlanguage}[1]{%
  \@ifundefined{alias@\string#1}
    {\ORIGselectlanguage{#1}}
    {\begingroup\edef\x{\endgroup
       \noexpand\ORIGselectlanguage{\@nameuse{alias@#1}}}\x}%
}
\newcommand{\definelanguagealias}[2]{%
  \@namedef{alias@#1}{#2}%
}

\makeatother

\definelanguagealias{en}{english}
\definelanguagealias{EN}{english}
\definelanguagealias{eng}{english}
\definelanguagealias{de}{ngerman}

\begin{document}

\title{Dynamical spectra from one and two-photon Fock state pulses exciting a single chiral qubit in a waveguide}

\author{Sofia Arranz Regidor}
\email{18sar4@queensu.ca}
\affiliation{Department of Physics,
Engineering Physics and Astronomy, Queen's University, Kingston, Ontario, Canada, K7L 3N6}
\author{Andreas Knorr}
\affiliation{Institut für Theoretische Physik, Nichtlineare Optik und Quantenelektronik, Technische Universität Berlin, Berlin, 10623, Germany}
\author{Stephen Hughes}
\affiliation{Department of Physics,
Engineering Physics and Astronomy, Queen's University, Kingston, Ontario, Canada, K7L 3N6}
\email{shughes@queensu.ca}

\date{\today}

\begin{abstract} 
We study the dynamical light emission from few-photon Fock states in waveguide-QED with a chiral two-level system. 
We first investigate the time dynamics of the system by calculating the emitter population and illustrate the breakdown of the weak excitation approximation, for both 1-photon and 2-photon excitation.
We show how a 1-photon pulse yields a transmitted long-time spectrum that is identical to the input pulse, despite significant population effects. However, the dynamical spectra and spectral intensity show rich population effects.
We also show the differences between 1-photon and 2-photon excitation,
where the latter shows 
clear 
signatures
of nonlinear saturation effects. 
Analytical  and  
 numerically exact matrix product state solutions are shown.
\end{abstract}

\maketitle

{\it Introduction.}---The study of waveguide quantum electrodynamics (QED)  has significantly improved our ability to control quantum light-matter interactions on chip, where quantum 
bits represented by  
two-level systems (TLSs) are coupled to a continuum of quantized field modes~\cite{Witthaut_2010,PhysRevLett.106.053601,PhysRevLett.113.263604,Nysteen2015,PhysRevA.83.063828,Longhi:21,Hoi2015,PhysRevLett.116.093601,PhysRevA.82.063816,Kannan2020,Kannan2023,Lodahl2017,RevModPhys.89.021001,PhysRevLett.123.233602,PhysRevLett.110.013601,PhysRevA.101.023807,PhysRevA.106.013714,PhysRevA.110.L031703}. 
Waveguide photons coupled to atoms and resonant few-state systems present a fundamental model in quantum optics, and serve as a platform for
emerging  quantum optical 
applications ~\cite{PhysRevLett.115.153901,PhysRevLett.101.113903,PhysRevResearch.4.023082,Mirhosseini2019,RevModPhys.95.015002,PhysRevLett.115.163603,PhysRevX.2.011014,PhysRevLett.131.103602,PhysRevA.92.063836,PhysRevX.7.031024,PhysRevLett.120.140404}.

\begin{figure}[t]
\centering
\includegraphics[width=0.99\columnwidth]{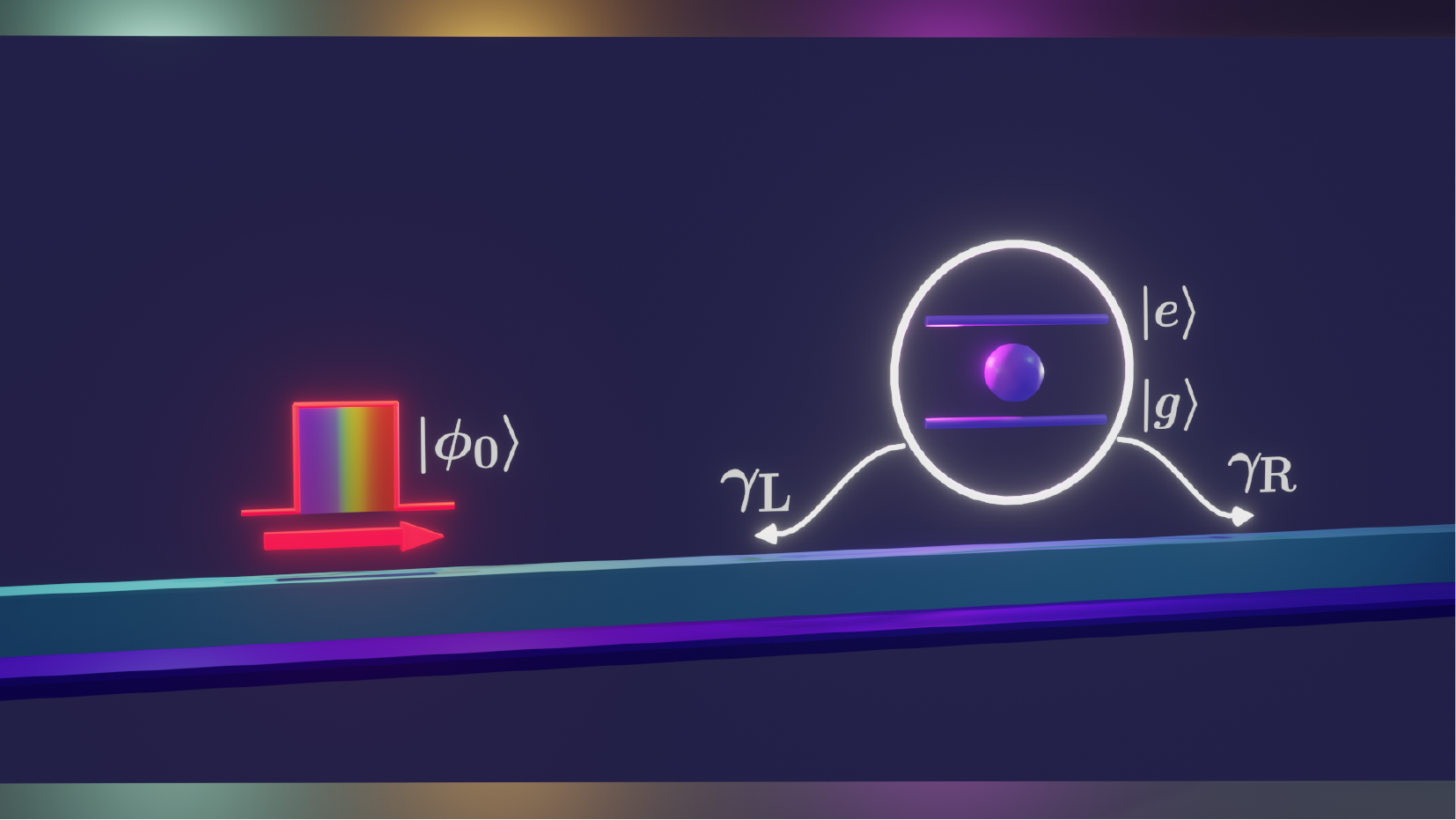}
\caption{Schematic of a TLS coupled to an open waveguide, excited with an incident rectangular quantum pulse, 
which can be a 1-photon or 2-photon 
 Fock (number) state.}
\label{fig:schem}
\end{figure}

Modelling quantized fields, for instance few photon states from waveguides interacting with TLSs presents a significant challenge, and often
one adopts various approximations.
For example, one can 
  treat the waveguide as a 
 {\it bath} that is subsequently traced out, or consider the system in a steady state (long time limit), or/and using a {\it weak excitation approximation} (WEA) where the TLSs 
 remain unexcited.
Often the WEA yields 
 identical solutions to a quantized harmonic oscillator (HO)~\cite{PhysRevA.79.023837,PhysRevResearch.2.043213,PhysRevLett.121.143601}, but even at the single photon limit when involving finite electron population, the WEA can fundamentally breakdown. 
There are established methods to model
the  exact few-photon transport in waveguide-QED, such as scattering solutions and input-output theories~\cite{PhysRevA.31.3761,PhysRevLett.98.153003,PhysRevA.76.062709,Rephaeli2012FewPhotonSC,PhysRevA.82.063821},
which are convenient for computing
long-time solutions.
More direct dynamical approaches include
time-dependent wavefunctions~\cite{Chen_2011,Nysteen2015},
and 
Heisenberg treatments~\cite{PhysRevA.106.023708}.

In this work, we study a
 quantum pulse containing only a few photons (one or two) interacting with a single chiral TLS in a waveguide (Fig.~\ref{fig:schem}). 
To fully include the
TLS dynamics and waveguide photons
at the system level, we use matrix product states (MPS), an exact numerical method based on tensor network methods, which allows us to 
treat the TLS and the waveguide at the same level~\cite{PhysRevResearch.3.023030,PhysRevA.103.033704,SnchezBurillo2015,Guimond_2017,PhysRevX.10.031011}. 
We complement our numerical approach by deriving 
analytical solutions for the TLS population for both a 1-photon and a 2-photon pulse.
We also use MPS to obtain the time-dependent correlation functions and dynamical (time-dependent) spectra~\cite{PhysRevA.106.023708,PhysRevA.65.033832,PhysRevA.66.063810}, and we derive the analytical solution of the correlation functions for the one-photon solution. 

Based on this approach we gain several key insights: 
First, we investigate the time dynamics of the waveguide-TLS system, observing the breakdown of the WEA and clear TLS
population dynamics, even for a single 
photon (as is well known~\cite{PhysRevA.82.033804}). 
Then we examine the transmitted spectra and observe that, strikingly, for 1-photon pulses,
the TLS population effects are completely cancelled 
out in the long time limit for the transmitted spectrum.
So how does one detect signatures of the population effects?
We show how the time-dependent spectra and spectral intensity 
do  contain  signatures of 
the TLS population effects,
showing significantly richer features than the usual 
CW resonance fluorescence features or the simple input spectrum of the pulse. 
For the 2-photon case, both the time-dependent spectra
and the stationary (long time) spectra contain unique saturation effects, and show a significant spectral narrowing
or reshaping.
These results introduce a way to probe the intrinsic population and nonlinear effects
that can occur in short-pulse dynamical resonance 
fluorescence, and are timely with recent experiments
in this regime, done so far with classical coherent fields~\cite{Liu2024,PhysRevLett.132.053602}.

{\it Theory.---}We consider a chiral TLS 
that scatters to the right only, 
so $\gamma_L=0$ and $\gamma=\gamma_R$ is the nominal decay rate. Symmetric emitter solutions 
are shown in Ref.~\onlinecite{sofia2024}.
Using the 
equations of motion
for the probability of the time-dependent atomic excitation amplitude of the TLS,  
we calculate the TLS population analytically~\cite{PhysRevA.106.023708,Chen_2011}.
We consider the frequency center of the photon wave packet pulse and TLS  on resonance,
$\omega_p=\omega_0$, with the 
group velocity  $v_g$ (linear dispersion).
We consider a square wave pulse shape, of width
$t_p$.
For the 1-photon (Fock state) pulse, we obtain
the TLS population dynamics~\cite{sofia2024}:
\begin{equation}
\begin{split}
    n_{\rm TLS}^{(1)} (t) =  {\frac{4}{\gamma t_p}} \left[e^{-\gamma t/2} -1   
    \right]^2, \qquad &0 \leq t \leq t_p, \\
    n_{\rm TLS}^{(1)} (t) =  {\frac{4}{\gamma t_p}} \left[ 1 
    - e^{\gamma t_p}
    \right]^2 e^{-\gamma t}, \qquad &t>t_p,
    \label{eq1}
\end{split}
\end{equation}
and for the 2-photon pulse,
\begin{equation}
\begin{split}
    n_{\rm TLS}^{(2)} (t) &=  {\frac{8}{\gamma t_p}} \Bigg[ \left( \frac{-32}{\gamma t_p} + \frac{16 t}{t_p} - 2 \right) e^{-\gamma t/2} -\frac{8}{\gamma t_p} + 1\\
    &+ \left( \frac{40}{\gamma t_p} + \frac{8 t}{t_p}  +1 \right) e^{-\gamma t}  
    \Bigg],
    \qquad 0 \leq t \leq t_p, \\
    n_{\rm TLS}^{(2)} (t) &=  \frac{8}{\gamma t_p} \Bigg[ \left(  \frac{-32}{\gamma t_p} +14 \right) e^{\gamma t_p /2} + \left(  \frac{-8}{\gamma t_p} +1 \right) e^{\gamma t_p } \\
    &+ \left(  \frac{40}{\gamma t_p} +9 \right) 
    \Bigg] e^{-\gamma t},
    \qquad t>t_p.
\end{split}
\end{equation}
To obtain the full dynamical information for the quantum correlation functions
and spectra, 
we use MPS~\cite{PhysRevResearch.3.023030}.

The temporal shape of any general quantum pulse can be defined in the input creation operator $b_{\rm in}^\dagger$, which creates the photons in the waveguide,
$b_{\rm in}^\dagger = \int dt \ f(t) b_R(t)^\dagger$,
where $f(t)$ represents the pulse shape and $b_R(t)$ is the photon creation operator that creates a right moving photon at a time $t$.
In the MPS description, this is discretized and introduced in the bosonic part of the input state. In the case of a single photon pulse,
then
\begin{equation}
    \ket{\phi_0}_N = \sum_{k=1}^N f_k \Delta B_k^\dagger /\sqrt{\Delta t} \ket{0,...,0},
\end{equation}
where $f_k$ is the discretized version of $f(t)$, and $\Delta B_k^\dagger = \int_{t_k}^{t_{k+1}} dt' b_R^\dagger(t')$ is the quantum noise operator~\cite{sofia2024}.

\begin{figure}[t]
\centering
\includegraphics[width=\columnwidth]{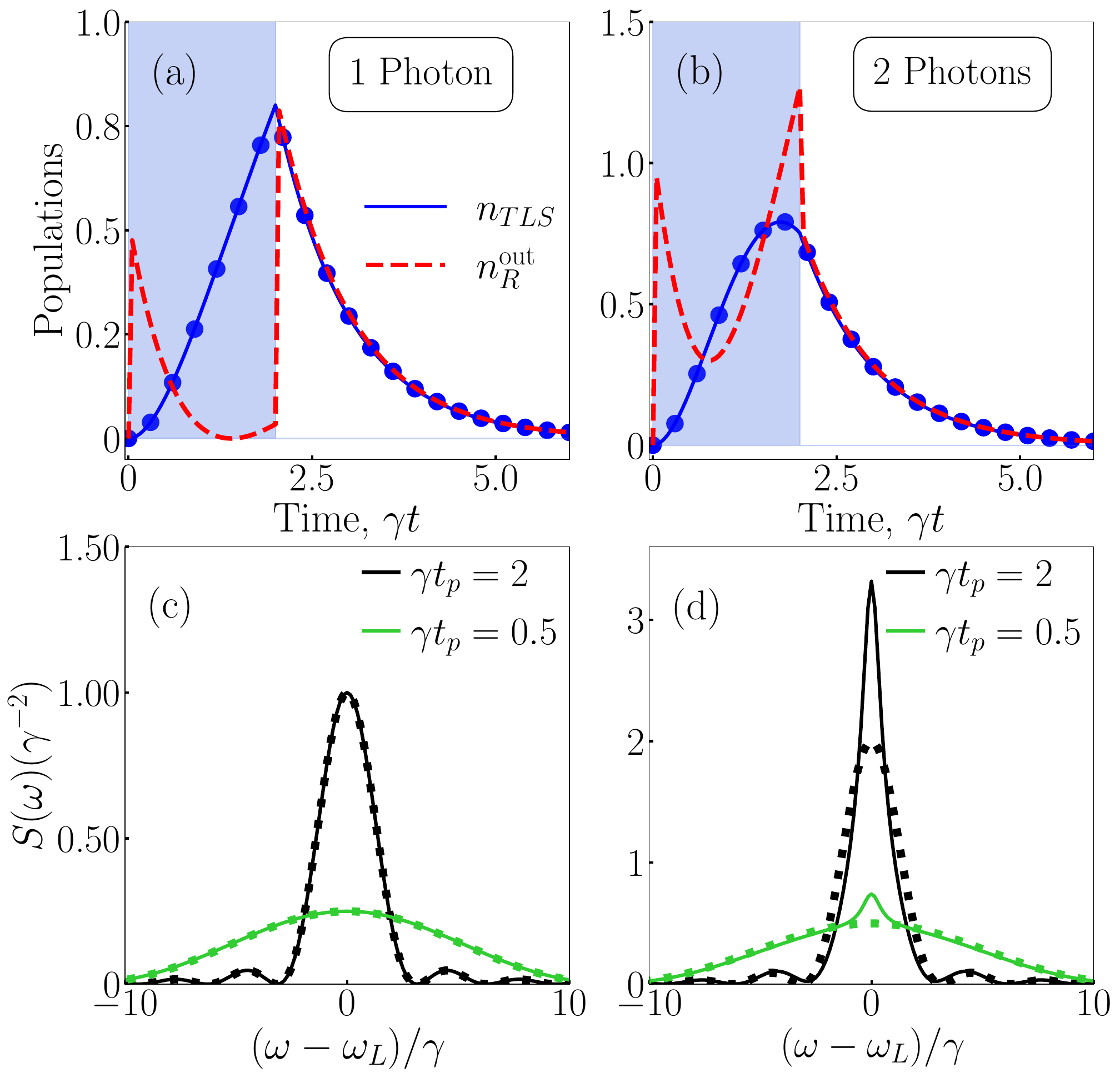}
\caption{(a,b) Quantum dynamics for a chiral single TLS-waveguide system, using an incident rectangular quantum pulse (with duration 2$\gamma^{-1}$, cyan shaded region) containing: (a) one photon, and (b) two photons. Both panels show the TLS population (solid blue), with the corresponding analytical solution (blue circles), and the normalized photon population of the right moving photons (dashed red).
 (c)Stationary spectrum, $S(\omega)$, for 1-photon with $\gamma t_p=2$ (black solid line) and $\gamma t_p=0.5$ (green solid line), and the corresponding solutions without the TLS interaction (dashed lines). 
 (d)  $S(\omega)$ for a 2-photon pulse, with similar labels
 to (c), where 
 dashed lines show solutions
 without the TLS interaction. 
 }
\label{fig:2}
\end{figure}

{\it Population dynamics and transmitted stationary spectra.---}In Fig.~\ref{fig:2}, we study an incident rectangular pulse with a width of $ \gamma t_p = 2$ (shaded region), containing either 1 photon 
[Fig~\ref{fig:2}(a)] or 2 photons [Fig.~\ref{fig:2}(b)], interacting with the chiral TLS. The population dynamics 
are calculated analytically (blue circles) and using MPS (solid line), which show perfect agreement. The normalized photon population
$
n_{R}^{\rm out} =  \braket{\Delta B_k^\dagger  \Delta B_k} /\left(\Delta t\right)^2
$
is shown as well for the right-moving 
photons (dashed red). 

For these relatively short pulses, 
we observe a high 
TLS population as the pulse interacts with the TLS, which then exponentially decays. These results completely deviate from the weak excitation regime, even in the 1-photon case, consistent with~\cite{PhysRevA.82.033804}. Furthermore, the TLS population is higher in the single photon case. This is due to a nonlinear feature in the 2-photon pulse solution, highlighting the complex nonlinear population saturation interactions with two photons. These population effects are further studied in Ref.~\onlinecite{sofia2024}, where the effects of pulse duration are investigated, showing significant TLS populations, greater than 0.1 for pulses as long as $100 \gamma$.

Figure~\ref{fig:2}(c) shows the 
input and transmitted 1-photon stationary spectrum for the same pulse duration as in (a) ($\gamma t_p=2$) and for a shorter pulse duration of $\gamma t_p =0.5$. For both pulse durations, all the population effects we observed in the TLS dynamical figures 
perfectly cancel out, i.e., the spectrum is insensitive to population dynamics. This is why a WEA solution 
can happen to give the
same result as a full TLS solution with dynamical excitation, even though a WEA assumption for the population is clearly not valid.
However, this 
correspondence is not the case when the input pulse contains 2 photons,
as shown in Fig~\ref{fig:2}(d), where the transmitted spectra is calculated again for a pulse durations of $\gamma t_p=2$ and $\gamma t_p=0.5$, and compared with the input pulse. Here, a clear nonlinear feature can be observed, where an additional peak with bandwidth $\gamma$ appears at the central frequency, qualitatively differing from the input field spectra.

\begin{figure*}[t]
\centering
\includegraphics[width=\textwidth]{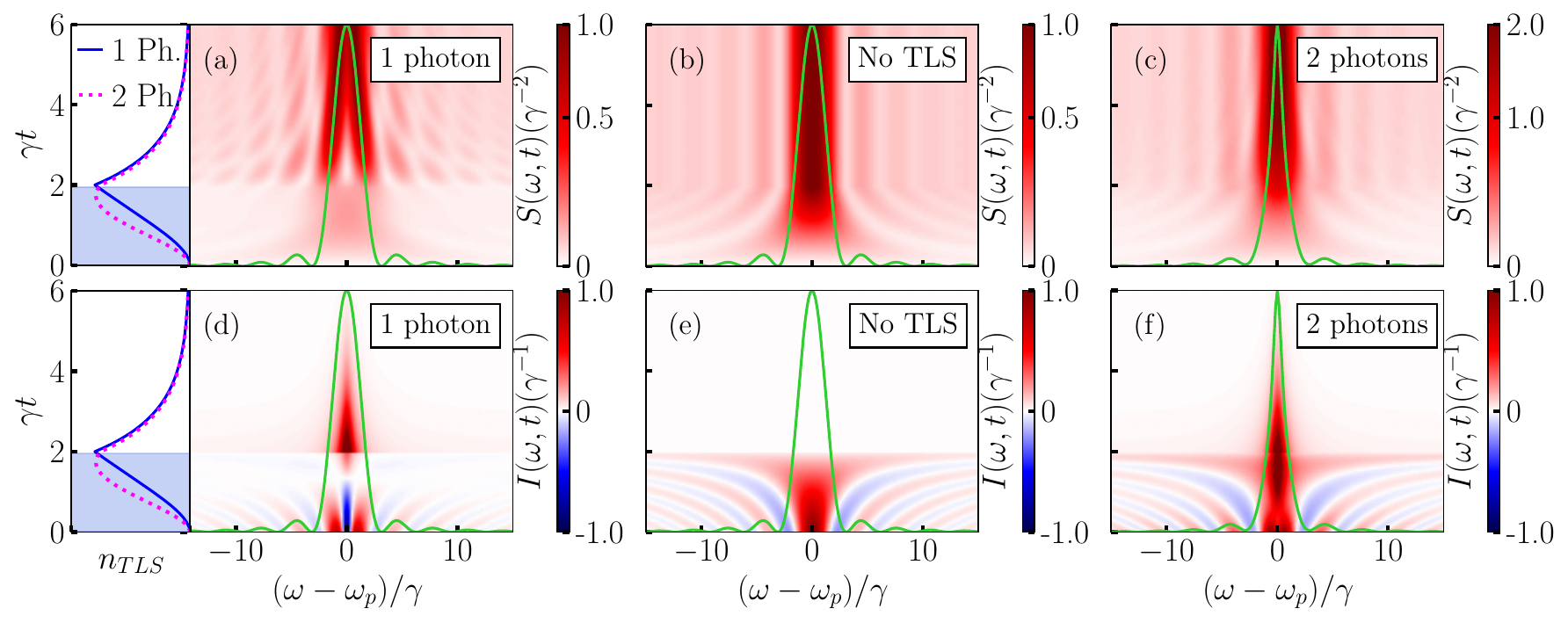}
\caption{(a,b,c) Time-dependent spectrum, $S(\omega,t)$, of (a) a 1-photon and (b) a 2-photon pulse with a pulse length $t_p=2/\gamma$ interacting with a chiral TLS, and (c) a 1-photon pulse of the same length without the TLS interaction. (d,e,f) Time-dependent spectral intensity, $I(\omega,t)$, of (d) a 1-photon and (e) a 2-photon quantum pulse with a pulse length $t_p=2/\gamma$ interacting with a chiral TLS, and (f) a 1-photon pulse of the same length without the TLS interaction. The corresponding long time limit $S(\omega)$, is plotted in the six cases in green. On the left, the TLS population ($n_{TLS}$) is plotted for both the interaction with a 1-photon pulse (solid blue curve) and a 2-photon pulse (dashed magenta curve) and the pulse is shown with a blue shade. 
}
\label{fig:4}
\end{figure*}

{\it Analytical solution
for the 1-photon long-time 
spectra.---}In Fig.~\ref{fig:2}(c), we 
 used MPS to compute the
stationary spectra,
and find that the transmitted spectrum is identical to the input spectra, even though
Fig.~\ref{fig:2}(a)
shows significant population effects.
To help explain the 
absence of TLS population effects
for the long-time  spectra, in the case of 1-photon excitation, 
we can use the analytical methods 
in Ref.~\onlinecite{sofia2024}.

We 
consider a right-propagating 1-photon field $a_R(t)$,
described through the boson operator, 
$a_{\rm in}(t)=a_0 f(t)/\sqrt{v_g}$ 
containing the pulse shape $f(t)$ and the annihilation operator $a_0$ that annihilates a waveguide photon  $a_0 \ket{0} =0$.
The transmitted spectrum $S(\omega)$ is
\begin{equation}
S(\omega) = v_g
\left< \left( a_{\rm in}^\dagger(\omega) -\frac{g_{0}^R}{v_g} \sigma^+(\omega) \right) \left( a_{\rm in}(\omega) -\frac{g_{0}^R}{v_g} \sigma^-(\omega) \right) \right>
\label{spectrum},
\end{equation}
where
$\braket{ a_{\rm in}^\dagger(\omega)a_{\rm in}(\omega)} = |f(\omega)|^2$, $\sigma^{+/-}(\omega)$ are the raising and lowering operators for the TLS, and $g^R_0$ is the (right travelling) photon-TLS coupling term.
Solving Eq.~\eqref{spectrum}
explicitly 
(full derivation in~\cite{sofia2024}), yields
  \begin{equation}
    \begin{split}
        S(\omega) & = |f(\omega)|^2 - \frac{\gamma \left| f(\omega)\right|^2}{i(\omega -\delta)+\gamma}  
        - \frac{\gamma \left| f(\omega)\right|^2}{-i(\omega -\delta)+\gamma}  
        \\ &
        + \frac{ \gamma^2 \left| f(\omega)\right|^2}{(\omega-\delta)^2+\gamma^2} 
         \equiv |f(\omega)|^2 ,
    \end{split} 
    \label{spectrumsol}
    \end{equation}
where $\delta=\omega_a-\omega_0$, with $\omega_a$ the TLS frequency.
Thus, we simply obtain the incident pulse profile, and all TLS effects perfectly cancel out, in agreement with results shown in Fig.~\ref{fig:2}(c). Interestingly, Eq.~\eqref{spectrumsol} is valid for any pulse shape despite the fact that different $f(t)$ generate different pulse dynamics.

{\it Dynamical spectra.---}To better understand the time-dependent population dynamics and how it manifests in experiments, we define and connect to two
dynamical observables that have also been probed in recent experiments \cite{Liu2024,PhysRevLett.132.053602}:
time-dependent (dynamical) spectra,
$S(\omega,t)$ and 
frequency-dependent 
 intensity, $I(\omega,t)$. 
In a rotating wave approximation, the transmitted {\it time-dependent spectrum} is obtained from
\begin{equation}
	S(\omega,t) = \text{Re}\left[ \int_0^{t} dt' \int_0^{t-t'} \!\! d\tau v_g \langle a_R^{\dagger}(t')a_R(t'+\tau)\rangle e^{i\Delta_{\omega {\rm p}}\tau} \right], 
 \label{eq:Swt}
\end{equation}
where $\Delta_{\omega {\rm p}} = \omega - \omega_{\rm p}$.
In the long time limit, when $t\rightarrow \infty$, we 
obtain $S(\omega) = S(\omega,t{\rightarrow} \infty)$. 
We stress that $S(\omega,t)$ contains features that are unique to the regime of 
{\it dynamical resonance fluorescence}.

For the same regime of pulsed excitations, we can also 
compute the emitted
time-dependent spectral intensity,
\begin{equation}
 I(\omega, t) = 
     {\rm Re} \left[ \int_0^{\infty} \! d\tau  v_g \braket{a_R^{\dagger} (t) a_R(t+\tau)}  e^{i \Delta _{\rm \omega {\rm p}}\tau}    \right],
    \label{eqI}
\end{equation}
and we note 
 the interesting relationship:
$S(\omega) = I(\omega)  = \int_0^{\infty}  I(\omega, t) dt $~\cite{Liu2024}, so the time-integrated spectral intensity is equivalent to the long-time spectra, but the dynamical signatures are quite different.
We highlight that the time-dependent spectral intensity was recently measured in 
cavity-QED experiments with {\it classical} pulses~\cite{Liu2024}, specifically with quantum dots in cavities, using 
an etalon filter and a fast 
photon detector.

Figure~\ref{fig:4} shows the time-dependent spectra, $S(\omega,t)$, calculated using Eq.~\eqref{eq:Swt} for a pulse of length $\gamma t_p=2$ containing 1 photon [Fig.~\ref{fig:4}(a)]  and 2 photons 
[Fig.~\ref{fig:4}(c)].
Figure \ref{fig:4}(b) represents the solution without the TLS for reference, which has naturally no population effects; 
note that neglecting population effects
in the dynamical case leads to unphysical time-dependent spectra,
$S(\omega,t)$,
(which is negative), but identical stationary spectra.
In addition, the TLS population is shown on the left to better understand the correlation with the population dynamics.
{\it In this dynamical case, we can appreciate the time-dependent TLS population effects on the spectrum, even in the 1-photon solution, both during the pulse and after the pulse. }
Additionally, we observe a more pronounced center peak in the 2-photon solution (compared to the 1-photon case),
manifested by nonlinear  effects.

\begin{figure}[t]
\centering
\includegraphics[width=\columnwidth]{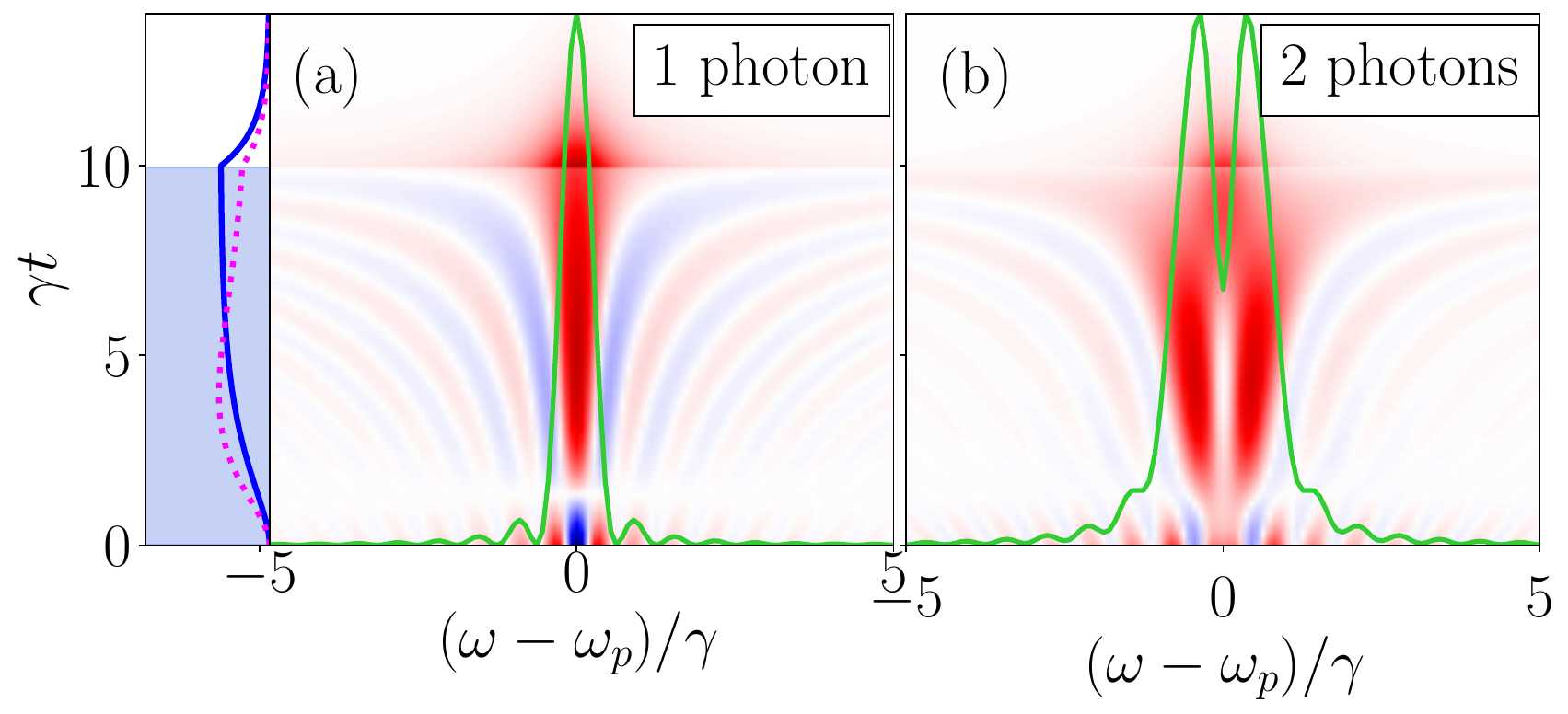}
\caption{ Time-dependent spectral intensity, $I(\omega,t)$, of (a) a 1-photon and (b) a 2-photon quantum pulse with a pulse length $t_p=10/\gamma$ interacting with a chiral TLS. The corresponding stationary spectra $S(\omega)$, is plotted in green. On the left, the TLS population ($n_{TLS}$) is plotted for both the interaction with a 1-photon pulse (solid blue curve) and a 2-photon pulse (dashed magenta curve) and the pulse is shown with a blue shade.}
\label{fig:5}
\end{figure}

The frequency-dependent spectral intensity [Eq.~\eqref{eqI}] results are shown in 
Fig.~\ref{fig:4}(d,e,f), where again we show the interaction of the TLS with a 1-photon pulse, the reference pulse, and the interaction with a 2-photon pulse, respectively.
Here, we observe a qualitatively different profile, as a function of time, as we now also probe part of the dynamics of the two-time photon correlation function. For instance, we can have positive and negative values, though in practice, actual measurements will yield a net positive quantity when using spectral filters~\cite{Liu2024}.
Nevertheless, we again see clear TLS population effects in both the 1 and 2-photon solutions, with an enhancement of the central peak in the latter case. 
This is again observed in Fig.~\ref{fig:5}, with a longer pulse ($\gamma t_p = 10$), where both results show temporal dynamics. Now, we see how the 1-photon case has more subtle dynamics since the TLS population is smaller than with shorter pulses. However, in the 2-photon results, there is a clear nonlinear interaction showing a pulse break-up in frequency,
consistent with results on collective 
chiral emitters~\cite{PhysRevX.10.031011}.

We also see how the time-integrated intensity, $I(\omega)$, coincides with the stationary spectrum, $S(\omega)$ (solid green lines in Figs.~\ref{fig:4} and~\ref{fig:5}), where the TLS population effects are perfectly cancelled for the 1-photon pulse.  
Again, this feature is only true in the long time limit of Eq.~\eqref{eq:Swt}; in contrast,
the {\it time-dependent} spectra and intensity instead contain clear signatures of population effects even for 
1 photon excitation.
For 2 photon excitation, we also obtain nonlinear effects, both in the time-dependent spectral functions as well as the stationary spectra.

To gain further insight into the role
of the population effects,
we have derived an analytical solution for the relevant two-time correlation 
function (full derivation can be found in Ref.~\cite{sofia2024}). For example,
when $t>t_p$, we have
    \begin{equation}
      v_g\braket{ a_R^\dagger(t)  a_R(t+\tau)} 
      = \frac{4}{t_p} \left( e^{\gamma t_p/2} -1 \right)^2 e^{-\gamma (t +\tau/2)},
    \end{equation}
    which uses the single-time solution as the initial condition, from
    ($t>t_p$)
    \begin{equation}
    v_g \braket{ a_R^\dagger(t)  a_R(t)} 
     = \frac{4}{ t_p} \left[ e^{\gamma t_p/2} -1 \right]^2 e^{-\gamma t},
    \end{equation}
    which closely resembles the population dynamics (see Eq.~\eqref{eq1}), 
    showing that population effects directly affect both $S(\omega,t)$
    and $I(\omega,t)$.
Thus, our results suggest that dynamical spectra can be used 
to measure dynamical TLS population effects with few photon
sources.

{\em Summary.---}We have investigated the time dynamics of a chiral TLS in a waveguide when excited with a rectangular quantum pulse containing few photon excitations, specifically using an $n$-photon Fock state. We have shown how the WEA breaks down,
and significant population features are observed with both 1- and 2-photon pulses. 
For the case of 1 photon excitation,
we have demonstrated how the TLS dynamical signatures cannot be observed in the stationary spectrum, since all the population effects cancel out in the long time limit.  
However, the impact of the population dynamics
is readily shown in the
dynamical 
spectral functions, 
for both $S(\omega,t)$ and $I(\omega,t)$,
computed 
 using a numerically exact  MPS approach. These spectral observables represent dynamical solutions which notably 
 contain population effects
 in the time-dependent spectrum solution.
 For 2-photon excitation, 
 we showed how both 
 the stationary and 
 dynamical spectra contain rich
 population and nonlinear features.

Apart from fundamental quantum optics interest,
our results also impact many current models in pulsed waveguide-QED systems with few photon sources, as the WEA is often adopted in many of the scattering theories;
however, 
our predictions do not rely on any weak excitation assumptions,
and our results are also timely with recent experimental works on measuring 
the {\it dynamical} resonance florescence regime~\cite{Liu2024,PhysRevLett.132.053602}.
Finally,
while our results were presented for a short rectangular pulse, 
our general conclusions and models can be extended to longer pulses, and for any quantum 
pulse shape, including Gaussian time profiles~\cite{sofia2024}. 
Indeed, we find that significant population effects occur 
even for relatively long pulses, with populations  $N_{TLS}>0.05$ for
durations $t_p  > 100/\gamma$~\cite{sofia2024}.

This work was supported by the Natural Sciences and Engineering Research Council of Canada (NSERC) [Discovery and Quantum Alliance Grants],
the National Research Council of Canada (NRC),
the Canadian Foundation for Innovation (CFI), Queen's University, Canada, and the 
Alexander von Humboldt Foundation 
through a Humboldt Award. 
We also thank Kisa Barkemeyer,
Jacob Ewaniuk, and Nir Rotenberg
for valuable contributions and discussions.

\bibliography{references_all}

\begin{thebibliography}{50}%
\makeatletter
\providecommand \@ifxundefined [1]{%
 \@ifx{#1\undefined}
}%
\providecommand \@ifnum [1]{%
 \ifnum #1\expandafter \@firstoftwo
 \else \expandafter \@secondoftwo
 \fi
}%
\providecommand \@ifx [1]{%
 \ifx #1\expandafter \@firstoftwo
 \else \expandafter \@secondoftwo
 \fi
}%
\providecommand \natexlab [1]{#1}%
\providecommand \enquote  [1]{``#1''}%
\providecommand \bibnamefont  [1]{#1}%
\providecommand \bibfnamefont [1]{#1}%
\providecommand \citenamefont [1]{#1}%
\providecommand \href@noop [0]{\@secondoftwo}%
\providecommand \href [0]{\begingroup \@sanitize@url \@href}%
\providecommand \@href[1]{\@@startlink{#1}\@@href}%
\providecommand \@@href[1]{\endgroup#1\@@endlink}%
\providecommand \@sanitize@url [0]{\catcode `\\12\catcode `\$12\catcode
  `\&12\catcode `\#12\catcode `\^12\catcode `\_12\catcode `\%12\relax}%
\providecommand \@@startlink[1]{}%
\providecommand \@@endlink[0]{}%
\providecommand \url  [0]{\begingroup\@sanitize@url \@url }%
\providecommand \@url [1]{\endgroup\@href {#1}{\urlprefix }}%
\providecommand \urlprefix  [0]{URL }%
\providecommand \Eprint [0]{\href }%
\providecommand \doibase [0]{https://doi.org/}%
\providecommand \selectlanguage [0]{\@gobble}%
\providecommand \bibinfo  [0]{\@secondoftwo}%
\providecommand \bibfield  [0]{\@secondoftwo}%
\providecommand \translation [1]{[#1]}%
\providecommand \BibitemOpen [0]{}%
\providecommand \bibitemStop [0]{}%
\providecommand \bibitemNoStop [0]{.\EOS\space}%
\providecommand \EOS [0]{\spacefactor3000\relax}%
\providecommand \BibitemShut  [1]{\csname bibitem#1\endcsname}%
\let\auto@bib@innerbib\@empty
\bibitem [{\citenamefont {Witthaut}\ and\ \citenamefont
  {Sørensen}(2010)}]{Witthaut_2010}%
  \BibitemOpen
  \bibfield  {author} {\bibinfo {author} {\bibfnamefont {D.}~\bibnamefont
  {Witthaut}}\ and\ \bibinfo {author} {\bibfnamefont {A.~S.}\ \bibnamefont
  {Sørensen}},\ }\bibfield  {title} {\bibinfo {title} {Photon scattering by a
  three-level emitter in a one-dimensional waveguide},\ }\href
  {https://doi.org/10.1088/1367-2630/12/4/043052} {\bibfield  {journal}
  {\bibinfo  {journal} {New Journal of Physics}\ }\textbf {\bibinfo {volume}
  {12}},\ \bibinfo {pages} {043052} (\bibinfo {year} {2010})}\BibitemShut
  {NoStop}%
\bibitem [{\citenamefont {Roy}(2011)}]{PhysRevLett.106.053601}%
  \BibitemOpen
  \bibfield  {author} {\bibinfo {author} {\bibfnamefont {D.}~\bibnamefont
  {Roy}},\ }\bibfield  {title} {\bibinfo {title} {Two-photon scattering by a
  driven three-level emitter in a one-dimensional waveguide and
  electromagnetically induced transparency},\ }\href
  {https://doi.org/10.1103/PhysRevLett.106.053601} {\bibfield  {journal}
  {\bibinfo  {journal} {Phys. Rev. Lett.}\ }\textbf {\bibinfo {volume} {106}},\
  \bibinfo {pages} {053601} (\bibinfo {year} {2011})}\BibitemShut {NoStop}%
\bibitem [{\citenamefont {Sanchez-Burillo}\ \emph {et~al.}(2014)\citenamefont
  {Sanchez-Burillo}, \citenamefont {Zueco}, \citenamefont {Garcia-Ripoll},\
  and\ \citenamefont {Martin-Moreno}}]{PhysRevLett.113.263604}%
  \BibitemOpen
  \bibfield  {author} {\bibinfo {author} {\bibfnamefont {E.}~\bibnamefont
  {Sanchez-Burillo}}, \bibinfo {author} {\bibfnamefont {D.}~\bibnamefont
  {Zueco}}, \bibinfo {author} {\bibfnamefont {J.~J.}\ \bibnamefont
  {Garcia-Ripoll}},\ and\ \bibinfo {author} {\bibfnamefont {L.}~\bibnamefont
  {Martin-Moreno}},\ }\bibfield  {title} {\bibinfo {title} {Scattering in the
  ultrastrong regime: Nonlinear optics with one photon},\ }\href
  {https://doi.org/10.1103/PhysRevLett.113.263604} {\bibfield  {journal}
  {\bibinfo  {journal} {Phys. Rev. Lett.}\ }\textbf {\bibinfo {volume} {113}},\
  \bibinfo {pages} {263604} (\bibinfo {year} {2014})}\BibitemShut {NoStop}%
\bibitem [{\citenamefont {Nysteen}\ \emph {et~al.}(2015)\citenamefont
  {Nysteen}, \citenamefont {Kristensen}, \citenamefont {McCutcheon},
  \citenamefont {Kaer},\ and\ \citenamefont {Mørk}}]{Nysteen2015}%
  \BibitemOpen
  \bibfield  {author} {\bibinfo {author} {\bibfnamefont {A.}~\bibnamefont
  {Nysteen}}, \bibinfo {author} {\bibfnamefont {P.~T.}\ \bibnamefont
  {Kristensen}}, \bibinfo {author} {\bibfnamefont {D.~P.~S.}\ \bibnamefont
  {McCutcheon}}, \bibinfo {author} {\bibfnamefont {P.}~\bibnamefont {Kaer}},\
  and\ \bibinfo {author} {\bibfnamefont {J.}~\bibnamefont {Mørk}},\ }\bibfield
   {title} {\bibinfo {title} {Scattering of two photons on a quantum emitter in
  a one-dimensional waveguide: exact dynamics and induced correlations},\
  }\href {https://doi.org/10.1088/1367-2630/17/2/023030} {\bibfield  {journal}
  {\bibinfo  {journal} {New Journal of Physics}\ }\textbf {\bibinfo {volume}
  {17}},\ \bibinfo {pages} {023030} (\bibinfo {year} {2015})}\BibitemShut
  {NoStop}%
\bibitem [{\citenamefont {Longo}\ \emph {et~al.}(2011)\citenamefont {Longo},
  \citenamefont {Schmitteckert},\ and\ \citenamefont
  {Busch}}]{PhysRevA.83.063828}%
  \BibitemOpen
  \bibfield  {author} {\bibinfo {author} {\bibfnamefont {P.}~\bibnamefont
  {Longo}}, \bibinfo {author} {\bibfnamefont {P.}~\bibnamefont
  {Schmitteckert}},\ and\ \bibinfo {author} {\bibfnamefont {K.}~\bibnamefont
  {Busch}},\ }\bibfield  {title} {\bibinfo {title} {Few-photon transport in
  low-dimensional systems},\ }\href
  {https://doi.org/10.1103/PhysRevA.83.063828} {\bibfield  {journal} {\bibinfo
  {journal} {Phys. Rev. A}\ }\textbf {\bibinfo {volume} {83}},\ \bibinfo
  {pages} {063828} (\bibinfo {year} {2011})}\BibitemShut {NoStop}%
\bibitem [{\citenamefont {Longhi}(2021)}]{Longhi:21}%
  \BibitemOpen
  \bibfield  {author} {\bibinfo {author} {\bibfnamefont {S.}~\bibnamefont
  {Longhi}},\ }\bibfield  {title} {\bibinfo {title} {Rabi oscillations of bound
  states in the continuum},\ }\href {https://doi.org/10.1364/OL.424756}
  {\bibfield  {journal} {\bibinfo  {journal} {Opt. Lett.}\ }\textbf {\bibinfo
  {volume} {46}},\ \bibinfo {pages} {2091} (\bibinfo {year}
  {2021})}\BibitemShut {NoStop}%
\bibitem [{\citenamefont {Hoi}\ \emph {et~al.}(2015)\citenamefont {Hoi},
  \citenamefont {Kockum}, \citenamefont {Tornberg}, \citenamefont
  {Pourkabirian}, \citenamefont {Johansson}, \citenamefont {Delsing},\ and\
  \citenamefont {Wilson}}]{Hoi2015}%
  \BibitemOpen
  \bibfield  {author} {\bibinfo {author} {\bibfnamefont {I.-C.}\ \bibnamefont
  {Hoi}}, \bibinfo {author} {\bibfnamefont {A.~F.}\ \bibnamefont {Kockum}},
  \bibinfo {author} {\bibfnamefont {L.}~\bibnamefont {Tornberg}}, \bibinfo
  {author} {\bibfnamefont {A.}~\bibnamefont {Pourkabirian}}, \bibinfo {author}
  {\bibfnamefont {G.}~\bibnamefont {Johansson}}, \bibinfo {author}
  {\bibfnamefont {P.}~\bibnamefont {Delsing}},\ and\ \bibinfo {author}
  {\bibfnamefont {C.~M.}\ \bibnamefont {Wilson}},\ }\bibfield  {title}
  {\bibinfo {title} {Probing the quantum vacuum with an artificial atom in
  front of a mirror},\ }\href {https://doi.org/10.1038/nphys3484} {\bibfield
  {journal} {\bibinfo  {journal} {Nature Physics}\ }\textbf {\bibinfo {volume}
  {11}},\ \bibinfo {pages} {1045–1049} (\bibinfo {year} {2015})}\BibitemShut
  {NoStop}%
\bibitem [{\citenamefont {Pichler}\ and\ \citenamefont
  {Zoller}(2016)}]{PhysRevLett.116.093601}%
  \BibitemOpen
  \bibfield  {author} {\bibinfo {author} {\bibfnamefont {H.}~\bibnamefont
  {Pichler}}\ and\ \bibinfo {author} {\bibfnamefont {P.}~\bibnamefont
  {Zoller}},\ }\bibfield  {title} {\bibinfo {title} {Photonic circuits with
  time delays and quantum feedback},\ }\href
  {https://doi.org/10.1103/PhysRevLett.116.093601} {\bibfield  {journal}
  {\bibinfo  {journal} {Phys. Rev. Lett.}\ }\textbf {\bibinfo {volume} {116}},\
  \bibinfo {pages} {093601} (\bibinfo {year} {2016})}\BibitemShut {NoStop}%
\bibitem [{\citenamefont {Zheng}\ \emph {et~al.}(2010)\citenamefont {Zheng},
  \citenamefont {Gauthier},\ and\ \citenamefont
  {Baranger}}]{PhysRevA.82.063816}%
  \BibitemOpen
  \bibfield  {author} {\bibinfo {author} {\bibfnamefont {H.}~\bibnamefont
  {Zheng}}, \bibinfo {author} {\bibfnamefont {D.~J.}\ \bibnamefont
  {Gauthier}},\ and\ \bibinfo {author} {\bibfnamefont {H.~U.}\ \bibnamefont
  {Baranger}},\ }\bibfield  {title} {\bibinfo {title} {Waveguide {QED}:
  Many-body bound-state effects in coherent and {Fock-state} scattering from a
  two-level system},\ }\href {https://doi.org/10.1103/PhysRevA.82.063816}
  {\bibfield  {journal} {\bibinfo  {journal} {Phys. Rev. A}\ }\textbf {\bibinfo
  {volume} {82}},\ \bibinfo {pages} {063816} (\bibinfo {year}
  {2010})}\BibitemShut {NoStop}%
\bibitem [{\citenamefont {Kannan}\ \emph {et~al.}(2020)\citenamefont {Kannan},
  \citenamefont {Ruckriegel}, \citenamefont {Campbell}, \citenamefont
  {Frisk~Kockum}, \citenamefont {Braum\"{u}ller}, \citenamefont {Kim},
  \citenamefont {Kjaergaard}, \citenamefont {Krantz}, \citenamefont {Melville},
  \citenamefont {Niedzielski}, \citenamefont {Veps\"{a}l\"{a}inen},
  \citenamefont {Winik}, \citenamefont {Yoder}, \citenamefont {Nori},
  \citenamefont {Orlando}, \citenamefont {Gustavsson},\ and\ \citenamefont
  {Oliver}}]{Kannan2020}%
  \BibitemOpen
  \bibfield  {author} {\bibinfo {author} {\bibfnamefont {B.}~\bibnamefont
  {Kannan}}, \bibinfo {author} {\bibfnamefont {M.~J.}\ \bibnamefont
  {Ruckriegel}}, \bibinfo {author} {\bibfnamefont {D.~L.}\ \bibnamefont
  {Campbell}}, \bibinfo {author} {\bibfnamefont {A.}~\bibnamefont
  {Frisk~Kockum}}, \bibinfo {author} {\bibfnamefont {J.}~\bibnamefont
  {Braum\"{u}ller}}, \bibinfo {author} {\bibfnamefont {D.~K.}\ \bibnamefont
  {Kim}}, \bibinfo {author} {\bibfnamefont {M.}~\bibnamefont {Kjaergaard}},
  \bibinfo {author} {\bibfnamefont {P.}~\bibnamefont {Krantz}}, \bibinfo
  {author} {\bibfnamefont {A.}~\bibnamefont {Melville}}, \bibinfo {author}
  {\bibfnamefont {B.~M.}\ \bibnamefont {Niedzielski}}, \bibinfo {author}
  {\bibfnamefont {A.}~\bibnamefont {Veps\"{a}l\"{a}inen}}, \bibinfo {author}
  {\bibfnamefont {R.}~\bibnamefont {Winik}}, \bibinfo {author} {\bibfnamefont
  {J.~L.}\ \bibnamefont {Yoder}}, \bibinfo {author} {\bibfnamefont
  {F.}~\bibnamefont {Nori}}, \bibinfo {author} {\bibfnamefont {T.~P.}\
  \bibnamefont {Orlando}}, \bibinfo {author} {\bibfnamefont {S.}~\bibnamefont
  {Gustavsson}},\ and\ \bibinfo {author} {\bibfnamefont {W.~D.}\ \bibnamefont
  {Oliver}},\ }\bibfield  {title} {\bibinfo {title} {Waveguide quantum
  electrodynamics with superconducting artificial giant atoms},\ }\href
  {https://doi.org/10.1038/s41586-020-2529-9} {\bibfield  {journal} {\bibinfo
  {journal} {Nature}\ }\textbf {\bibinfo {volume} {583}},\ \bibinfo {pages}
  {775–779} (\bibinfo {year} {2020})}\BibitemShut {NoStop}%
\bibitem [{\citenamefont {Kannan}\ \emph {et~al.}(2023)\citenamefont {Kannan},
  \citenamefont {Almanakly}, \citenamefont {Sung}, \citenamefont {Di~Paolo},
  \citenamefont {Rower}, \citenamefont {Braum\"{u}ller}, \citenamefont
  {Melville}, \citenamefont {Niedzielski}, \citenamefont {Karamlou},
  \citenamefont {Serniak}, \citenamefont {Veps\"{a}l\"{a}inen}, \citenamefont
  {Schwartz}, \citenamefont {Yoder}, \citenamefont {Winik}, \citenamefont
  {Wang}, \citenamefont {Orlando}, \citenamefont {Gustavsson}, \citenamefont
  {Grover},\ and\ \citenamefont {Oliver}}]{Kannan2023}%
  \BibitemOpen
  \bibfield  {author} {\bibinfo {author} {\bibfnamefont {B.}~\bibnamefont
  {Kannan}}, \bibinfo {author} {\bibfnamefont {A.}~\bibnamefont {Almanakly}},
  \bibinfo {author} {\bibfnamefont {Y.}~\bibnamefont {Sung}}, \bibinfo {author}
  {\bibfnamefont {A.}~\bibnamefont {Di~Paolo}}, \bibinfo {author}
  {\bibfnamefont {D.~A.}\ \bibnamefont {Rower}}, \bibinfo {author}
  {\bibfnamefont {J.}~\bibnamefont {Braum\"{u}ller}}, \bibinfo {author}
  {\bibfnamefont {A.}~\bibnamefont {Melville}}, \bibinfo {author}
  {\bibfnamefont {B.~M.}\ \bibnamefont {Niedzielski}}, \bibinfo {author}
  {\bibfnamefont {A.}~\bibnamefont {Karamlou}}, \bibinfo {author}
  {\bibfnamefont {K.}~\bibnamefont {Serniak}}, \bibinfo {author} {\bibfnamefont
  {A.}~\bibnamefont {Veps\"{a}l\"{a}inen}}, \bibinfo {author} {\bibfnamefont
  {M.~E.}\ \bibnamefont {Schwartz}}, \bibinfo {author} {\bibfnamefont {J.~L.}\
  \bibnamefont {Yoder}}, \bibinfo {author} {\bibfnamefont {R.}~\bibnamefont
  {Winik}}, \bibinfo {author} {\bibfnamefont {J.~I.-J.}\ \bibnamefont {Wang}},
  \bibinfo {author} {\bibfnamefont {T.~P.}\ \bibnamefont {Orlando}}, \bibinfo
  {author} {\bibfnamefont {S.}~\bibnamefont {Gustavsson}}, \bibinfo {author}
  {\bibfnamefont {J.~A.}\ \bibnamefont {Grover}},\ and\ \bibinfo {author}
  {\bibfnamefont {W.~D.}\ \bibnamefont {Oliver}},\ }\bibfield  {title}
  {\bibinfo {title} {On-demand directional microwave photon emission using
  waveguide quantum electrodynamics},\ }\href
  {https://doi.org/10.1038/s41567-022-01869-5} {\bibfield  {journal} {\bibinfo
  {journal} {Nature Physics}\ }\textbf {\bibinfo {volume} {19}},\ \bibinfo
  {pages} {394–400} (\bibinfo {year} {2023})}\BibitemShut {NoStop}%
\bibitem [{\citenamefont {Lodahl}\ \emph {et~al.}(2017)\citenamefont {Lodahl},
  \citenamefont {Mahmoodian}, \citenamefont {Stobbe}, \citenamefont
  {Rauschenbeutel}, \citenamefont {Schneeweiss}, \citenamefont {Volz},
  \citenamefont {Pichler},\ and\ \citenamefont {Zoller}}]{Lodahl2017}%
  \BibitemOpen
  \bibfield  {author} {\bibinfo {author} {\bibfnamefont {P.}~\bibnamefont
  {Lodahl}}, \bibinfo {author} {\bibfnamefont {S.}~\bibnamefont {Mahmoodian}},
  \bibinfo {author} {\bibfnamefont {S.}~\bibnamefont {Stobbe}}, \bibinfo
  {author} {\bibfnamefont {A.}~\bibnamefont {Rauschenbeutel}}, \bibinfo
  {author} {\bibfnamefont {P.}~\bibnamefont {Schneeweiss}}, \bibinfo {author}
  {\bibfnamefont {J.}~\bibnamefont {Volz}}, \bibinfo {author} {\bibfnamefont
  {H.}~\bibnamefont {Pichler}},\ and\ \bibinfo {author} {\bibfnamefont
  {P.}~\bibnamefont {Zoller}},\ }\bibfield  {title} {\bibinfo {title} {Chiral
  quantum optics},\ }\href {https://doi.org/10.1038/nature21037} {\bibfield
  {journal} {\bibinfo  {journal} {Nature}\ }\textbf {\bibinfo {volume} {541}},\
  \bibinfo {pages} {473} (\bibinfo {year} {2017})}\BibitemShut {NoStop}%
\bibitem [{\citenamefont {Roy}\ \emph {et~al.}(2017)\citenamefont {Roy},
  \citenamefont {Wilson},\ and\ \citenamefont
  {Firstenberg}}]{RevModPhys.89.021001}%
  \BibitemOpen
  \bibfield  {author} {\bibinfo {author} {\bibfnamefont {D.}~\bibnamefont
  {Roy}}, \bibinfo {author} {\bibfnamefont {C.~M.}\ \bibnamefont {Wilson}},\
  and\ \bibinfo {author} {\bibfnamefont {O.}~\bibnamefont {Firstenberg}},\
  }\bibfield  {title} {\bibinfo {title} {Colloquium: Strongly interacting
  photons in one-dimensional continuum},\ }\href
  {https://doi.org/10.1103/RevModPhys.89.021001} {\bibfield  {journal}
  {\bibinfo  {journal} {Rev. Mod. Phys.}\ }\textbf {\bibinfo {volume} {89}},\
  \bibinfo {pages} {021001} (\bibinfo {year} {2017})}\BibitemShut {NoStop}%
\bibitem [{\citenamefont {Wen}\ \emph {et~al.}(2019)\citenamefont {Wen},
  \citenamefont {Lin}, \citenamefont {Kockum}, \citenamefont {Suri},
  \citenamefont {Ian}, \citenamefont {Chen}, \citenamefont {Mao}, \citenamefont
  {Chiu}, \citenamefont {Delsing}, \citenamefont {Nori}, \citenamefont {Lin},\
  and\ \citenamefont {Hoi}}]{PhysRevLett.123.233602}%
  \BibitemOpen
  \bibfield  {author} {\bibinfo {author} {\bibfnamefont {P.~Y.}\ \bibnamefont
  {Wen}}, \bibinfo {author} {\bibfnamefont {K.-T.}\ \bibnamefont {Lin}},
  \bibinfo {author} {\bibfnamefont {A.~F.}\ \bibnamefont {Kockum}}, \bibinfo
  {author} {\bibfnamefont {B.}~\bibnamefont {Suri}}, \bibinfo {author}
  {\bibfnamefont {H.}~\bibnamefont {Ian}}, \bibinfo {author} {\bibfnamefont
  {J.~C.}\ \bibnamefont {Chen}}, \bibinfo {author} {\bibfnamefont {S.~Y.}\
  \bibnamefont {Mao}}, \bibinfo {author} {\bibfnamefont {C.~C.}\ \bibnamefont
  {Chiu}}, \bibinfo {author} {\bibfnamefont {P.}~\bibnamefont {Delsing}},
  \bibinfo {author} {\bibfnamefont {F.}~\bibnamefont {Nori}}, \bibinfo {author}
  {\bibfnamefont {G.-D.}\ \bibnamefont {Lin}},\ and\ \bibinfo {author}
  {\bibfnamefont {I.-C.}\ \bibnamefont {Hoi}},\ }\bibfield  {title} {\bibinfo
  {title} {Large collective {Lamb} shift of two distant superconducting
  artificial atoms},\ }\href {https://doi.org/10.1103/PhysRevLett.123.233602}
  {\bibfield  {journal} {\bibinfo  {journal} {Phys. Rev. Lett.}\ }\textbf
  {\bibinfo {volume} {123}},\ \bibinfo {pages} {233602} (\bibinfo {year}
  {2019})}\BibitemShut {NoStop}%
\bibitem [{\citenamefont {Carmele}\ \emph {et~al.}(2013)\citenamefont
  {Carmele}, \citenamefont {Kabuss}, \citenamefont {Schulze}, \citenamefont
  {Reitzenstein},\ and\ \citenamefont {Knorr}}]{PhysRevLett.110.013601}%
  \BibitemOpen
  \bibfield  {author} {\bibinfo {author} {\bibfnamefont {A.}~\bibnamefont
  {Carmele}}, \bibinfo {author} {\bibfnamefont {J.}~\bibnamefont {Kabuss}},
  \bibinfo {author} {\bibfnamefont {F.}~\bibnamefont {Schulze}}, \bibinfo
  {author} {\bibfnamefont {S.}~\bibnamefont {Reitzenstein}},\ and\ \bibinfo
  {author} {\bibfnamefont {A.}~\bibnamefont {Knorr}},\ }\bibfield  {title}
  {\bibinfo {title} {Single photon delayed feedback: A way to stabilize
  intrinsic quantum cavity electrodynamics},\ }\href
  {https://doi.org/10.1103/PhysRevLett.110.013601} {\bibfield  {journal}
  {\bibinfo  {journal} {Phys. Rev. Lett.}\ }\textbf {\bibinfo {volume} {110}},\
  \bibinfo {pages} {013601} (\bibinfo {year} {2013})}\BibitemShut {NoStop}%
\bibitem [{\citenamefont {Crowder}\ \emph {et~al.}(2020)\citenamefont
  {Crowder}, \citenamefont {Carmichael},\ and\ \citenamefont
  {Hughes}}]{PhysRevA.101.023807}%
  \BibitemOpen
  \bibfield  {author} {\bibinfo {author} {\bibfnamefont {G.}~\bibnamefont
  {Crowder}}, \bibinfo {author} {\bibfnamefont {J.~H.}\ \bibnamefont
  {Carmichael}},\ and\ \bibinfo {author} {\bibfnamefont {S.}~\bibnamefont
  {Hughes}},\ }\bibfield  {title} {\bibinfo {title} {Quantum trajectory theory
  of few-photon {cavity-{QED}} systems with a time-delayed coherent feedback},\
  }\href {https://doi.org/10.1103/PhysRevA.101.023807} {\bibfield  {journal}
  {\bibinfo  {journal} {Phys. Rev. A}\ }\textbf {\bibinfo {volume} {101}},\
  \bibinfo {pages} {023807} (\bibinfo {year} {2020})}\BibitemShut {NoStop}%
\bibitem [{\citenamefont {Crowder}\ \emph {et~al.}(2022)\citenamefont
  {Crowder}, \citenamefont {Ramunno},\ and\ \citenamefont
  {Hughes}}]{PhysRevA.106.013714}%
  \BibitemOpen
  \bibfield  {author} {\bibinfo {author} {\bibfnamefont {G.}~\bibnamefont
  {Crowder}}, \bibinfo {author} {\bibfnamefont {L.}~\bibnamefont {Ramunno}},\
  and\ \bibinfo {author} {\bibfnamefont {S.}~\bibnamefont {Hughes}},\
  }\bibfield  {title} {\bibinfo {title} {Quantum trajectory theory and
  simulations of nonlinear spectra and multiphoton effects in {waveguide-{QED}}
  systems with a time-delayed coherent feedback},\ }\href
  {https://doi.org/10.1103/PhysRevA.106.013714} {\bibfield  {journal} {\bibinfo
   {journal} {Phys. Rev. A}\ }\textbf {\bibinfo {volume} {106}},\ \bibinfo
  {pages} {013714} (\bibinfo {year} {2022})}\BibitemShut {NoStop}%
\bibitem [{\citenamefont {Crowder}\ \emph {et~al.}(2024)\citenamefont
  {Crowder}, \citenamefont {Ramunno},\ and\ \citenamefont
  {Hughes}}]{PhysRevA.110.L031703}%
  \BibitemOpen
  \bibfield  {author} {\bibinfo {author} {\bibfnamefont {G.}~\bibnamefont
  {Crowder}}, \bibinfo {author} {\bibfnamefont {L.}~\bibnamefont {Ramunno}},\
  and\ \bibinfo {author} {\bibfnamefont {S.}~\bibnamefont {Hughes}},\
  }\bibfield  {title} {\bibinfo {title} {Improving on-demand
  single-photon-source coherence and indistinguishability through a
  time-delayed coherent feedback},\ }\href
  {https://doi.org/10.1103/PhysRevA.110.L031703} {\bibfield  {journal}
  {\bibinfo  {journal} {Phys. Rev. A}\ }\textbf {\bibinfo {volume} {110}},\
  \bibinfo {pages} {L031703} (\bibinfo {year} {2024})}\BibitemShut {NoStop}%
\bibitem [{\citenamefont {Young}\ \emph {et~al.}(2015)\citenamefont {Young},
  \citenamefont {Thijssen}, \citenamefont {Beggs}, \citenamefont
  {Androvitsaneas}, \citenamefont {Kuipers}, \citenamefont {Rarity},
  \citenamefont {Hughes},\ and\ \citenamefont
  {Oulton}}]{PhysRevLett.115.153901}%
  \BibitemOpen
  \bibfield  {author} {\bibinfo {author} {\bibfnamefont {A.~B.}\ \bibnamefont
  {Young}}, \bibinfo {author} {\bibfnamefont {A.~C.~T.}\ \bibnamefont
  {Thijssen}}, \bibinfo {author} {\bibfnamefont {D.~M.}\ \bibnamefont {Beggs}},
  \bibinfo {author} {\bibfnamefont {P.}~\bibnamefont {Androvitsaneas}},
  \bibinfo {author} {\bibfnamefont {L.}~\bibnamefont {Kuipers}}, \bibinfo
  {author} {\bibfnamefont {J.~G.}\ \bibnamefont {Rarity}}, \bibinfo {author}
  {\bibfnamefont {S.}~\bibnamefont {Hughes}},\ and\ \bibinfo {author}
  {\bibfnamefont {R.}~\bibnamefont {Oulton}},\ }\bibfield  {title} {\bibinfo
  {title} {Polarization engineering in photonic crystal waveguides for
  spin-photon entanglers},\ }\href
  {https://doi.org/10.1103/PhysRevLett.115.153901} {\bibfield  {journal}
  {\bibinfo  {journal} {Phys. Rev. Lett.}\ }\textbf {\bibinfo {volume} {115}},\
  \bibinfo {pages} {153901} (\bibinfo {year} {2015})}\BibitemShut {NoStop}%
\bibitem [{\citenamefont {Lund-Hansen}\ \emph {et~al.}(2008)\citenamefont
  {Lund-Hansen}, \citenamefont {Stobbe}, \citenamefont {Julsgaard},
  \citenamefont {Thyrrestrup}, \citenamefont {S\"unner}, \citenamefont {Kamp},
  \citenamefont {Forchel},\ and\ \citenamefont
  {Lodahl}}]{PhysRevLett.101.113903}%
  \BibitemOpen
  \bibfield  {author} {\bibinfo {author} {\bibfnamefont {T.}~\bibnamefont
  {Lund-Hansen}}, \bibinfo {author} {\bibfnamefont {S.}~\bibnamefont {Stobbe}},
  \bibinfo {author} {\bibfnamefont {B.}~\bibnamefont {Julsgaard}}, \bibinfo
  {author} {\bibfnamefont {H.}~\bibnamefont {Thyrrestrup}}, \bibinfo {author}
  {\bibfnamefont {T.}~\bibnamefont {S\"unner}}, \bibinfo {author}
  {\bibfnamefont {M.}~\bibnamefont {Kamp}}, \bibinfo {author} {\bibfnamefont
  {A.}~\bibnamefont {Forchel}},\ and\ \bibinfo {author} {\bibfnamefont
  {P.}~\bibnamefont {Lodahl}},\ }\bibfield  {title} {\bibinfo {title}
  {Experimental realization of highly efficient broadband coupling of single
  quantum dots to a photonic crystal waveguide},\ }\href
  {https://doi.org/10.1103/PhysRevLett.101.113903} {\bibfield  {journal}
  {\bibinfo  {journal} {Phys. Rev. Lett.}\ }\textbf {\bibinfo {volume} {101}},\
  \bibinfo {pages} {113903} (\bibinfo {year} {2008})}\BibitemShut {NoStop}%
\bibitem [{\citenamefont {Hauff}\ \emph {et~al.}(2022)\citenamefont {Hauff},
  \citenamefont {Le~Jeannic}, \citenamefont {Lodahl}, \citenamefont {Hughes},\
  and\ \citenamefont {Rotenberg}}]{PhysRevResearch.4.023082}%
  \BibitemOpen
  \bibfield  {author} {\bibinfo {author} {\bibfnamefont {N.~V.}\ \bibnamefont
  {Hauff}}, \bibinfo {author} {\bibfnamefont {H.}~\bibnamefont {Le~Jeannic}},
  \bibinfo {author} {\bibfnamefont {P.}~\bibnamefont {Lodahl}}, \bibinfo
  {author} {\bibfnamefont {S.}~\bibnamefont {Hughes}},\ and\ \bibinfo {author}
  {\bibfnamefont {N.}~\bibnamefont {Rotenberg}},\ }\bibfield  {title} {\bibinfo
  {title} {Chiral quantum optics in broken-symmetry and topological photonic
  crystal waveguides},\ }\href
  {https://doi.org/10.1103/PhysRevResearch.4.023082} {\bibfield  {journal}
  {\bibinfo  {journal} {Phys. Rev. Res.}\ }\textbf {\bibinfo {volume} {4}},\
  \bibinfo {pages} {023082} (\bibinfo {year} {2022})}\BibitemShut {NoStop}%
\bibitem [{\citenamefont {Mirhosseini}\ \emph {et~al.}(2019)\citenamefont
  {Mirhosseini}, \citenamefont {Kim}, \citenamefont {Zhang}, \citenamefont
  {Sipahigil}, \citenamefont {Dieterle}, \citenamefont {Keller}, \citenamefont
  {Asenjo-Garcia}, \citenamefont {Chang},\ and\ \citenamefont
  {Painter}}]{Mirhosseini2019}%
  \BibitemOpen
  \bibfield  {author} {\bibinfo {author} {\bibfnamefont {M.}~\bibnamefont
  {Mirhosseini}}, \bibinfo {author} {\bibfnamefont {E.}~\bibnamefont {Kim}},
  \bibinfo {author} {\bibfnamefont {X.}~\bibnamefont {Zhang}}, \bibinfo
  {author} {\bibfnamefont {A.}~\bibnamefont {Sipahigil}}, \bibinfo {author}
  {\bibfnamefont {P.~B.}\ \bibnamefont {Dieterle}}, \bibinfo {author}
  {\bibfnamefont {A.~J.}\ \bibnamefont {Keller}}, \bibinfo {author}
  {\bibfnamefont {A.}~\bibnamefont {Asenjo-Garcia}}, \bibinfo {author}
  {\bibfnamefont {D.~E.}\ \bibnamefont {Chang}},\ and\ \bibinfo {author}
  {\bibfnamefont {O.}~\bibnamefont {Painter}},\ }\bibfield  {title} {\bibinfo
  {title} {Cavity quantum electrodynamics with atom-like mirrors},\ }\href
  {https://doi.org/10.1038/s41586-019-1196-1} {\bibfield  {journal} {\bibinfo
  {journal} {Nature}\ }\textbf {\bibinfo {volume} {569}},\ \bibinfo {pages}
  {692} (\bibinfo {year} {2019})}\BibitemShut {NoStop}%
\bibitem [{\citenamefont {Sheremet}\ \emph {et~al.}(2023)\citenamefont
  {Sheremet}, \citenamefont {Petrov}, \citenamefont {Iorsh}, \citenamefont
  {Poshakinskiy},\ and\ \citenamefont {Poddubny}}]{RevModPhys.95.015002}%
  \BibitemOpen
  \bibfield  {author} {\bibinfo {author} {\bibfnamefont {A.~S.}\ \bibnamefont
  {Sheremet}}, \bibinfo {author} {\bibfnamefont {M.~I.}\ \bibnamefont
  {Petrov}}, \bibinfo {author} {\bibfnamefont {I.~V.}\ \bibnamefont {Iorsh}},
  \bibinfo {author} {\bibfnamefont {A.~V.}\ \bibnamefont {Poshakinskiy}},\ and\
  \bibinfo {author} {\bibfnamefont {A.~N.}\ \bibnamefont {Poddubny}},\
  }\bibfield  {title} {\bibinfo {title} {Waveguide quantum electrodynamics:
  Collective radiance and photon-photon correlations},\ }\href
  {https://doi.org/10.1103/RevModPhys.95.015002} {\bibfield  {journal}
  {\bibinfo  {journal} {Rev. Mod. Phys.}\ }\textbf {\bibinfo {volume} {95}},\
  \bibinfo {pages} {015002} (\bibinfo {year} {2023})}\BibitemShut {NoStop}%
\bibitem [{\citenamefont {Gonz\'alez-Tudela}\ \emph {et~al.}(2015)\citenamefont
  {Gonz\'alez-Tudela}, \citenamefont {Paulisch}, \citenamefont {Chang},
  \citenamefont {Kimble},\ and\ \citenamefont
  {Cirac}}]{PhysRevLett.115.163603}%
  \BibitemOpen
  \bibfield  {author} {\bibinfo {author} {\bibfnamefont {A.}~\bibnamefont
  {Gonz\'alez-Tudela}}, \bibinfo {author} {\bibfnamefont {V.}~\bibnamefont
  {Paulisch}}, \bibinfo {author} {\bibfnamefont {D.~E.}\ \bibnamefont {Chang}},
  \bibinfo {author} {\bibfnamefont {H.~J.}\ \bibnamefont {Kimble}},\ and\
  \bibinfo {author} {\bibfnamefont {J.~I.}\ \bibnamefont {Cirac}},\ }\bibfield
  {title} {\bibinfo {title} {Deterministic generation of arbitrary photonic
  states assisted by dissipation},\ }\href
  {https://doi.org/10.1103/PhysRevLett.115.163603} {\bibfield  {journal}
  {\bibinfo  {journal} {Phys. Rev. Lett.}\ }\textbf {\bibinfo {volume} {115}},\
  \bibinfo {pages} {163603} (\bibinfo {year} {2015})}\BibitemShut {NoStop}%
\bibitem [{\citenamefont {Laucht}\ \emph {et~al.}(2012)\citenamefont {Laucht},
  \citenamefont {P\"utz}, \citenamefont {G\"unthner}, \citenamefont {Hauke},
  \citenamefont {Saive}, \citenamefont {Fr\'ed\'erick}, \citenamefont
  {Bichler}, \citenamefont {Amann}, \citenamefont {Holleitner}, \citenamefont
  {Kaniber},\ and\ \citenamefont {Finley}}]{PhysRevX.2.011014}%
  \BibitemOpen
  \bibfield  {author} {\bibinfo {author} {\bibfnamefont {A.}~\bibnamefont
  {Laucht}}, \bibinfo {author} {\bibfnamefont {S.}~\bibnamefont {P\"utz}},
  \bibinfo {author} {\bibfnamefont {T.}~\bibnamefont {G\"unthner}}, \bibinfo
  {author} {\bibfnamefont {N.}~\bibnamefont {Hauke}}, \bibinfo {author}
  {\bibfnamefont {R.}~\bibnamefont {Saive}}, \bibinfo {author} {\bibfnamefont
  {S.}~\bibnamefont {Fr\'ed\'erick}}, \bibinfo {author} {\bibfnamefont
  {M.}~\bibnamefont {Bichler}}, \bibinfo {author} {\bibfnamefont {M.-C.}\
  \bibnamefont {Amann}}, \bibinfo {author} {\bibfnamefont {A.~W.}\ \bibnamefont
  {Holleitner}}, \bibinfo {author} {\bibfnamefont {M.}~\bibnamefont
  {Kaniber}},\ and\ \bibinfo {author} {\bibfnamefont {J.~J.}\ \bibnamefont
  {Finley}},\ }\bibfield  {title} {\bibinfo {title} {A waveguide-coupled
  on-chip single-photon source},\ }\href
  {https://doi.org/10.1103/PhysRevX.2.011014} {\bibfield  {journal} {\bibinfo
  {journal} {Phys. Rev. X}\ }\textbf {\bibinfo {volume} {2}},\ \bibinfo {pages}
  {011014} (\bibinfo {year} {2012})}\BibitemShut {NoStop}%
\bibitem [{\citenamefont {Nie}\ \emph {et~al.}(2023)\citenamefont {Nie},
  \citenamefont {Shi}, \citenamefont {Liu},\ and\ \citenamefont
  {Nori}}]{PhysRevLett.131.103602}%
  \BibitemOpen
  \bibfield  {author} {\bibinfo {author} {\bibfnamefont {W.}~\bibnamefont
  {Nie}}, \bibinfo {author} {\bibfnamefont {T.}~\bibnamefont {Shi}}, \bibinfo
  {author} {\bibfnamefont {Y.-x.}\ \bibnamefont {Liu}},\ and\ \bibinfo {author}
  {\bibfnamefont {F.}~\bibnamefont {Nori}},\ }\bibfield  {title} {\bibinfo
  {title} {{Non-Hermitian} waveguide cavity {QED} with tunable atomic
  mirrors},\ }\href {https://doi.org/10.1103/PhysRevLett.131.103602} {\bibfield
   {journal} {\bibinfo  {journal} {Phys. Rev. Lett.}\ }\textbf {\bibinfo
  {volume} {131}},\ \bibinfo {pages} {103602} (\bibinfo {year}
  {2023})}\BibitemShut {NoStop}%
\bibitem [{\citenamefont {Li}\ and\ \citenamefont
  {Wei}(2015)}]{PhysRevA.92.063836}%
  \BibitemOpen
  \bibfield  {author} {\bibinfo {author} {\bibfnamefont {X.}~\bibnamefont
  {Li}}\ and\ \bibinfo {author} {\bibfnamefont {L.~F.}\ \bibnamefont {Wei}},\
  }\bibfield  {title} {\bibinfo {title} {Designable single-photon quantum
  routings with atomic mirrors},\ }\href
  {https://doi.org/10.1103/PhysRevA.92.063836} {\bibfield  {journal} {\bibinfo
  {journal} {Phys. Rev. A}\ }\textbf {\bibinfo {volume} {92}},\ \bibinfo
  {pages} {063836} (\bibinfo {year} {2015})}\BibitemShut {NoStop}%
\bibitem [{\citenamefont {Asenjo-Garcia}\ \emph {et~al.}(2017)\citenamefont
  {Asenjo-Garcia}, \citenamefont {Moreno-Cardoner}, \citenamefont {Albrecht},
  \citenamefont {Kimble},\ and\ \citenamefont {Chang}}]{PhysRevX.7.031024}%
  \BibitemOpen
  \bibfield  {author} {\bibinfo {author} {\bibfnamefont {A.}~\bibnamefont
  {Asenjo-Garcia}}, \bibinfo {author} {\bibfnamefont {M.}~\bibnamefont
  {Moreno-Cardoner}}, \bibinfo {author} {\bibfnamefont {A.}~\bibnamefont
  {Albrecht}}, \bibinfo {author} {\bibfnamefont {H.~J.}\ \bibnamefont
  {Kimble}},\ and\ \bibinfo {author} {\bibfnamefont {D.~E.}\ \bibnamefont
  {Chang}},\ }\bibfield  {title} {\bibinfo {title} {Exponential improvement in
  photon storage fidelities using subradiance and ``selective radiance'' in
  atomic arrays},\ }\href {https://doi.org/10.1103/PhysRevX.7.031024}
  {\bibfield  {journal} {\bibinfo  {journal} {Phys. Rev. X}\ }\textbf {\bibinfo
  {volume} {7}},\ \bibinfo {pages} {031024} (\bibinfo {year}
  {2017})}\BibitemShut {NoStop}%
\bibitem [{\citenamefont {Kockum}\ \emph {et~al.}(2018)\citenamefont {Kockum},
  \citenamefont {Johansson},\ and\ \citenamefont
  {Nori}}]{PhysRevLett.120.140404}%
  \BibitemOpen
  \bibfield  {author} {\bibinfo {author} {\bibfnamefont {A.~F.}\ \bibnamefont
  {Kockum}}, \bibinfo {author} {\bibfnamefont {G.}~\bibnamefont {Johansson}},\
  and\ \bibinfo {author} {\bibfnamefont {F.}~\bibnamefont {Nori}},\ }\bibfield
  {title} {\bibinfo {title} {Decoherence-free interaction between giant atoms
  in waveguide quantum electrodynamics},\ }\href
  {https://doi.org/10.1103/PhysRevLett.120.140404} {\bibfield  {journal}
  {\bibinfo  {journal} {Phys. Rev. Lett.}\ }\textbf {\bibinfo {volume} {120}},\
  \bibinfo {pages} {140404} (\bibinfo {year} {2018})}\BibitemShut {NoStop}%
\bibitem [{\citenamefont {Shen}\ and\ \citenamefont
  {Fan}(2009)}]{PhysRevA.79.023837}%
  \BibitemOpen
  \bibfield  {author} {\bibinfo {author} {\bibfnamefont {J.-T.}\ \bibnamefont
  {Shen}}\ and\ \bibinfo {author} {\bibfnamefont {S.}~\bibnamefont {Fan}},\
  }\bibfield  {title} {\bibinfo {title} {Theory of single-photon transport in a
  single-mode waveguide. i. coupling to a cavity containing a two-level atom},\
  }\href {https://doi.org/10.1103/PhysRevA.79.023837} {\bibfield  {journal}
  {\bibinfo  {journal} {Phys. Rev. A}\ }\textbf {\bibinfo {volume} {79}},\
  \bibinfo {pages} {023837} (\bibinfo {year} {2009})}\BibitemShut {NoStop}%
\bibitem [{\citenamefont {Masson}\ and\ \citenamefont
  {Asenjo-Garcia}(2020)}]{PhysRevResearch.2.043213}%
  \BibitemOpen
  \bibfield  {author} {\bibinfo {author} {\bibfnamefont {S.~J.}\ \bibnamefont
  {Masson}}\ and\ \bibinfo {author} {\bibfnamefont {A.}~\bibnamefont
  {Asenjo-Garcia}},\ }\bibfield  {title} {\bibinfo {title} {Atomic-waveguide
  quantum electrodynamics},\ }\href
  {https://doi.org/10.1103/PhysRevResearch.2.043213} {\bibfield  {journal}
  {\bibinfo  {journal} {Phys. Rev. Res.}\ }\textbf {\bibinfo {volume} {2}},\
  \bibinfo {pages} {043213} (\bibinfo {year} {2020})}\BibitemShut {NoStop}%
\bibitem [{\citenamefont {Mahmoodian}\ \emph {et~al.}(2018)\citenamefont
  {Mahmoodian}, \citenamefont {\ifmmode~\check{C}\else \v{C}\fi{}epulkovskis},
  \citenamefont {Das}, \citenamefont {Lodahl}, \citenamefont {Hammerer},\ and\
  \citenamefont {S\o{}rensen}}]{PhysRevLett.121.143601}%
  \BibitemOpen
  \bibfield  {author} {\bibinfo {author} {\bibfnamefont {S.}~\bibnamefont
  {Mahmoodian}}, \bibinfo {author} {\bibfnamefont {M.}~\bibnamefont
  {\ifmmode~\check{C}\else \v{C}\fi{}epulkovskis}}, \bibinfo {author}
  {\bibfnamefont {S.}~\bibnamefont {Das}}, \bibinfo {author} {\bibfnamefont
  {P.}~\bibnamefont {Lodahl}}, \bibinfo {author} {\bibfnamefont
  {K.}~\bibnamefont {Hammerer}},\ and\ \bibinfo {author} {\bibfnamefont
  {A.~S.}\ \bibnamefont {S\o{}rensen}},\ }\bibfield  {title} {\bibinfo {title}
  {Strongly correlated photon transport in waveguide quantum electrodynamics
  with weakly coupled emitters},\ }\href
  {https://doi.org/10.1103/PhysRevLett.121.143601} {\bibfield  {journal}
  {\bibinfo  {journal} {Phys. Rev. Lett.}\ }\textbf {\bibinfo {volume} {121}},\
  \bibinfo {pages} {143601} (\bibinfo {year} {2018})}\BibitemShut {NoStop}%
\bibitem [{\citenamefont {Gardiner}\ and\ \citenamefont
  {Collett}(1985)}]{PhysRevA.31.3761}%
  \BibitemOpen
  \bibfield  {author} {\bibinfo {author} {\bibfnamefont {C.~W.}\ \bibnamefont
  {Gardiner}}\ and\ \bibinfo {author} {\bibfnamefont {M.~J.}\ \bibnamefont
  {Collett}},\ }\bibfield  {title} {\bibinfo {title} {Input and output in
  damped quantum systems: Quantum stochastic differential equations and the
  master equation},\ }\href {https://doi.org/10.1103/PhysRevA.31.3761}
  {\bibfield  {journal} {\bibinfo  {journal} {Phys. Rev. A}\ }\textbf {\bibinfo
  {volume} {31}},\ \bibinfo {pages} {3761} (\bibinfo {year}
  {1985})}\BibitemShut {NoStop}%
\bibitem [{\citenamefont {Shen}\ and\ \citenamefont
  {Fan}(2007{\natexlab{a}})}]{PhysRevLett.98.153003}%
  \BibitemOpen
  \bibfield  {author} {\bibinfo {author} {\bibfnamefont {J.-T.}\ \bibnamefont
  {Shen}}\ and\ \bibinfo {author} {\bibfnamefont {S.}~\bibnamefont {Fan}},\
  }\bibfield  {title} {\bibinfo {title} {Strongly correlated two-photon
  transport in a one-dimensional waveguide coupled to a two-level system},\
  }\href {https://doi.org/10.1103/PhysRevLett.98.153003} {\bibfield  {journal}
  {\bibinfo  {journal} {Phys. Rev. Lett.}\ }\textbf {\bibinfo {volume} {98}},\
  \bibinfo {pages} {153003} (\bibinfo {year} {2007}{\natexlab{a}})}\BibitemShut
  {NoStop}%
\bibitem [{\citenamefont {Shen}\ and\ \citenamefont
  {Fan}(2007{\natexlab{b}})}]{PhysRevA.76.062709}%
  \BibitemOpen
  \bibfield  {author} {\bibinfo {author} {\bibfnamefont {J.-T.}\ \bibnamefont
  {Shen}}\ and\ \bibinfo {author} {\bibfnamefont {S.}~\bibnamefont {Fan}},\
  }\bibfield  {title} {\bibinfo {title} {Strongly correlated multiparticle
  transport in one dimension through a quantum impurity},\ }\href
  {https://doi.org/10.1103/PhysRevA.76.062709} {\bibfield  {journal} {\bibinfo
  {journal} {Phys. Rev. A}\ }\textbf {\bibinfo {volume} {76}},\ \bibinfo
  {pages} {062709} (\bibinfo {year} {2007}{\natexlab{b}})}\BibitemShut
  {NoStop}%
\bibitem [{\citenamefont {Rephaeli}\ and\ \citenamefont
  {Fan}(2012)}]{Rephaeli2012FewPhotonSC}%
  \BibitemOpen
  \bibfield  {author} {\bibinfo {author} {\bibfnamefont {E.}~\bibnamefont
  {Rephaeli}}\ and\ \bibinfo {author} {\bibfnamefont {S.}~\bibnamefont {Fan}},\
  }\bibfield  {title} {\bibinfo {title} {Few-photon single-atom cavity {QED}
  with input-output formalism in fock space},\ }\href
  {https://api.semanticscholar.org/CorpusID:40913017} {\bibfield  {journal}
  {\bibinfo  {journal} {IEEE Journal of Selected Topics in Quantum
  Electronics}\ }\textbf {\bibinfo {volume} {18}},\ \bibinfo {pages} {1754}
  (\bibinfo {year} {2012})}\BibitemShut {NoStop}%
\bibitem [{\citenamefont {Fan}\ \emph {et~al.}(2010)\citenamefont {Fan},
  \citenamefont {Kocaba\mbox{\c{s}}},\ and\ \citenamefont
  {Shen}}]{PhysRevA.82.063821}%
  \BibitemOpen
  \bibfield  {author} {\bibinfo {author} {\bibfnamefont {S.}~\bibnamefont
  {Fan}}, \bibinfo {author} {\bibfnamefont {S.~E.}\ \bibnamefont
  {Kocaba\mbox{\c{s}}}},\ and\ \bibinfo {author} {\bibfnamefont {J.-T.}\
  \bibnamefont {Shen}},\ }\bibfield  {title} {\bibinfo {title} {Input-output
  formalism for few-photon transport in one-dimensional nanophotonic waveguides
  coupled to a qubit},\ }\href {https://doi.org/10.1103/PhysRevA.82.063821}
  {\bibfield  {journal} {\bibinfo  {journal} {Phys. Rev. A}\ }\textbf {\bibinfo
  {volume} {82}},\ \bibinfo {pages} {063821} (\bibinfo {year}
  {2010})}\BibitemShut {NoStop}%
\bibitem [{\citenamefont {Chen}\ \emph {et~al.}(2011)\citenamefont {Chen},
  \citenamefont {Wubs}, \citenamefont {Mørk},\ and\ \citenamefont
  {Koenderink}}]{Chen_2011}%
  \BibitemOpen
  \bibfield  {author} {\bibinfo {author} {\bibfnamefont {Y.}~\bibnamefont
  {Chen}}, \bibinfo {author} {\bibfnamefont {M.}~\bibnamefont {Wubs}}, \bibinfo
  {author} {\bibfnamefont {J.}~\bibnamefont {Mørk}},\ and\ \bibinfo {author}
  {\bibfnamefont {A.~F.}\ \bibnamefont {Koenderink}},\ }\bibfield  {title}
  {\bibinfo {title} {Coherent single-photon absorption by single emitters
  coupled to one-dimensional nanophotonic waveguides},\ }\href
  {https://doi.org/10.1088/1367-2630/13/10/103010} {\bibfield  {journal}
  {\bibinfo  {journal} {New Journal of Physics}\ }\textbf {\bibinfo {volume}
  {13}},\ \bibinfo {pages} {103010} (\bibinfo {year} {2011})}\BibitemShut
  {NoStop}%
\bibitem [{\citenamefont {Barkemeyer}\ \emph {et~al.}(2022)\citenamefont
  {Barkemeyer}, \citenamefont {Knorr},\ and\ \citenamefont
  {Carmele}}]{PhysRevA.106.023708}%
  \BibitemOpen
  \bibfield  {author} {\bibinfo {author} {\bibfnamefont {K.}~\bibnamefont
  {Barkemeyer}}, \bibinfo {author} {\bibfnamefont {A.}~\bibnamefont {Knorr}},\
  and\ \bibinfo {author} {\bibfnamefont {A.}~\bibnamefont {Carmele}},\
  }\bibfield  {title} {\bibinfo {title} {Heisenberg treatment of multiphoton
  pulses in waveguide {QED} with time-delayed feedback},\ }\href
  {https://doi.org/10.1103/PhysRevA.106.023708} {\bibfield  {journal} {\bibinfo
   {journal} {Phys. Rev. A}\ }\textbf {\bibinfo {volume} {106}},\ \bibinfo
  {pages} {023708} (\bibinfo {year} {2022})}\BibitemShut {NoStop}%
\bibitem [{\citenamefont {Arranz~Regidor}\ \emph {et~al.}(2021)\citenamefont
  {Arranz~Regidor}, \citenamefont {Crowder}, \citenamefont {Carmichael},\ and\
  \citenamefont {Hughes}}]{PhysRevResearch.3.023030}%
  \BibitemOpen
  \bibfield  {author} {\bibinfo {author} {\bibfnamefont {S.}~\bibnamefont
  {Arranz~Regidor}}, \bibinfo {author} {\bibfnamefont {G.}~\bibnamefont
  {Crowder}}, \bibinfo {author} {\bibfnamefont {H.}~\bibnamefont
  {Carmichael}},\ and\ \bibinfo {author} {\bibfnamefont {S.}~\bibnamefont
  {Hughes}},\ }\bibfield  {title} {\bibinfo {title} {Modeling quantum
  light-matter interactions in waveguide {QED} with retardation, nonlinear
  interactions, and a time-delayed feedback: Matrix product states versus a
  space-discretized waveguide model},\ }\href
  {https://doi.org/10.1103/PhysRevResearch.3.023030} {\bibfield  {journal}
  {\bibinfo  {journal} {Phys. Rev. Res.}\ }\textbf {\bibinfo {volume} {3}},\
  \bibinfo {pages} {023030} (\bibinfo {year} {2021})}\BibitemShut {NoStop}%
\bibitem [{\citenamefont {Barkemeyer}\ \emph {et~al.}(2021)\citenamefont
  {Barkemeyer}, \citenamefont {Knorr},\ and\ \citenamefont
  {Carmele}}]{PhysRevA.103.033704}%
  \BibitemOpen
  \bibfield  {author} {\bibinfo {author} {\bibfnamefont {K.}~\bibnamefont
  {Barkemeyer}}, \bibinfo {author} {\bibfnamefont {A.}~\bibnamefont {Knorr}},\
  and\ \bibinfo {author} {\bibfnamefont {A.}~\bibnamefont {Carmele}},\
  }\bibfield  {title} {\bibinfo {title} {Strongly entangled system-reservoir
  dynamics with multiphoton pulses beyond the two-excitation limit: Exciting
  the atom-photon bound state},\ }\href
  {https://doi.org/10.1103/PhysRevA.103.033704} {\bibfield  {journal} {\bibinfo
   {journal} {Phys. Rev. A}\ }\textbf {\bibinfo {volume} {103}},\ \bibinfo
  {pages} {033704} (\bibinfo {year} {2021})}\BibitemShut {NoStop}%
\bibitem [{\citenamefont {Sánchez-Burillo}\ \emph {et~al.}(2015)\citenamefont
  {Sánchez-Burillo}, \citenamefont {García-Ripoll}, \citenamefont
  {Martín-Moreno},\ and\ \citenamefont {Zueco}}]{SnchezBurillo2015}%
  \BibitemOpen
  \bibfield  {author} {\bibinfo {author} {\bibfnamefont {E.}~\bibnamefont
  {Sánchez-Burillo}}, \bibinfo {author} {\bibfnamefont {J.}~\bibnamefont
  {García-Ripoll}}, \bibinfo {author} {\bibfnamefont {L.}~\bibnamefont
  {Martín-Moreno}},\ and\ \bibinfo {author} {\bibfnamefont {D.}~\bibnamefont
  {Zueco}},\ }\bibfield  {title} {\bibinfo {title} {Nonlinear quantum optics in
  the (ultra)strong light–matter coupling},\ }\href
  {https://doi.org/10.1039/c4fd00206g} {\bibfield  {journal} {\bibinfo
  {journal} {Faraday Discussions}\ }\textbf {\bibinfo {volume} {178}},\
  \bibinfo {pages} {335–356} (\bibinfo {year} {2015})}\BibitemShut {NoStop}%
\bibitem [{\citenamefont {Guimond}\ \emph {et~al.}(2017)\citenamefont
  {Guimond}, \citenamefont {Pletyukhov}, \citenamefont {Pichler},\ and\
  \citenamefont {Zoller}}]{Guimond_2017}%
  \BibitemOpen
  \bibfield  {author} {\bibinfo {author} {\bibfnamefont {P.-O.}\ \bibnamefont
  {Guimond}}, \bibinfo {author} {\bibfnamefont {M.}~\bibnamefont {Pletyukhov}},
  \bibinfo {author} {\bibfnamefont {H.}~\bibnamefont {Pichler}},\ and\ \bibinfo
  {author} {\bibfnamefont {P.}~\bibnamefont {Zoller}},\ }\bibfield  {title}
  {\bibinfo {title} {Delayed coherent quantum feedback from a scattering theory
  and a matrix product state perspective},\ }\href
  {https://doi.org/10.1088/2058-9565/aa7f03} {\bibfield  {journal} {\bibinfo
  {journal} {Quantum Science and Technology}\ }\textbf {\bibinfo {volume}
  {2}},\ \bibinfo {pages} {044012} (\bibinfo {year} {2017})}\BibitemShut
  {NoStop}%
\bibitem [{\citenamefont {Mahmoodian}\ \emph {et~al.}(2020)\citenamefont
  {Mahmoodian}, \citenamefont {Calaj\'o}, \citenamefont {Chang}, \citenamefont
  {Hammerer},\ and\ \citenamefont {S\o{}rensen}}]{PhysRevX.10.031011}%
  \BibitemOpen
  \bibfield  {author} {\bibinfo {author} {\bibfnamefont {S.}~\bibnamefont
  {Mahmoodian}}, \bibinfo {author} {\bibfnamefont {G.}~\bibnamefont
  {Calaj\'o}}, \bibinfo {author} {\bibfnamefont {D.~E.}\ \bibnamefont {Chang}},
  \bibinfo {author} {\bibfnamefont {K.}~\bibnamefont {Hammerer}},\ and\
  \bibinfo {author} {\bibfnamefont {A.~S.}\ \bibnamefont {S\o{}rensen}},\
  }\bibfield  {title} {\bibinfo {title} {Dynamics of many-body photon bound
  states in chiral waveguide {QED}},\ }\href
  {https://doi.org/10.1103/PhysRevX.10.031011} {\bibfield  {journal} {\bibinfo
  {journal} {Phys. Rev. X}\ }\textbf {\bibinfo {volume} {10}},\ \bibinfo
  {pages} {031011} (\bibinfo {year} {2020})}\BibitemShut {NoStop}%
\bibitem [{\citenamefont {Domokos}\ \emph {et~al.}(2002)\citenamefont
  {Domokos}, \citenamefont {Horak},\ and\ \citenamefont
  {Ritsch}}]{PhysRevA.65.033832}%
  \BibitemOpen
  \bibfield  {author} {\bibinfo {author} {\bibfnamefont {P.}~\bibnamefont
  {Domokos}}, \bibinfo {author} {\bibfnamefont {P.}~\bibnamefont {Horak}},\
  and\ \bibinfo {author} {\bibfnamefont {H.}~\bibnamefont {Ritsch}},\
  }\bibfield  {title} {\bibinfo {title} {Quantum description of light-pulse
  scattering on a single atom in waveguides},\ }\href
  {https://doi.org/10.1103/PhysRevA.65.033832} {\bibfield  {journal} {\bibinfo
  {journal} {Phys. Rev. A}\ }\textbf {\bibinfo {volume} {65}},\ \bibinfo
  {pages} {033832} (\bibinfo {year} {2002})}\BibitemShut {NoStop}%
\bibitem [{\citenamefont {Dung}\ \emph {et~al.}(2002)\citenamefont {Dung},
  \citenamefont {Kn\"oll},\ and\ \citenamefont {Welsch}}]{PhysRevA.66.063810}%
  \BibitemOpen
  \bibfield  {author} {\bibinfo {author} {\bibfnamefont {H.~T.}\ \bibnamefont
  {Dung}}, \bibinfo {author} {\bibfnamefont {L.}~\bibnamefont {Kn\"oll}},\ and\
  \bibinfo {author} {\bibfnamefont {D.-G.}\ \bibnamefont {Welsch}},\ }\bibfield
   {title} {\bibinfo {title} {Resonant dipole-dipole interaction in the
  presence of dispersing and absorbing surroundings},\ }\href
  {https://doi.org/10.1103/PhysRevA.66.063810} {\bibfield  {journal} {\bibinfo
  {journal} {Phys. Rev. A}\ }\textbf {\bibinfo {volume} {66}},\ \bibinfo
  {pages} {063810} (\bibinfo {year} {2002})}\BibitemShut {NoStop}%
\bibitem [{\citenamefont {Rephaeli}\ \emph {et~al.}(2010)\citenamefont
  {Rephaeli}, \citenamefont {Shen},\ and\ \citenamefont
  {Fan}}]{PhysRevA.82.033804}%
  \BibitemOpen
  \bibfield  {author} {\bibinfo {author} {\bibfnamefont {E.}~\bibnamefont
  {Rephaeli}}, \bibinfo {author} {\bibfnamefont {J.-T.}\ \bibnamefont {Shen}},\
  and\ \bibinfo {author} {\bibfnamefont {S.}~\bibnamefont {Fan}},\ }\bibfield
  {title} {\bibinfo {title} {Full inversion of a two-level atom with a
  single-photon pulse in one-dimensional geometries},\ }\href
  {https://doi.org/10.1103/PhysRevA.82.033804} {\bibfield  {journal} {\bibinfo
  {journal} {Phys. Rev. A}\ }\textbf {\bibinfo {volume} {82}},\ \bibinfo
  {pages} {033804} (\bibinfo {year} {2010})}\BibitemShut {NoStop}%
\bibitem [{\citenamefont {Liu}\ \emph {et~al.}(2024)\citenamefont {Liu},
  \citenamefont {Gustin}, \citenamefont {Liu}, \citenamefont {Li},
  \citenamefont {Yu}, \citenamefont {Ni}, \citenamefont {Niu}, \citenamefont
  {Hughes}, \citenamefont {Wang},\ and\ \citenamefont {Liu}}]{Liu2024}%
  \BibitemOpen
  \bibfield  {author} {\bibinfo {author} {\bibfnamefont {S.}~\bibnamefont
  {Liu}}, \bibinfo {author} {\bibfnamefont {C.}~\bibnamefont {Gustin}},
  \bibinfo {author} {\bibfnamefont {H.}~\bibnamefont {Liu}}, \bibinfo {author}
  {\bibfnamefont {X.}~\bibnamefont {Li}}, \bibinfo {author} {\bibfnamefont
  {Y.}~\bibnamefont {Yu}}, \bibinfo {author} {\bibfnamefont {H.}~\bibnamefont
  {Ni}}, \bibinfo {author} {\bibfnamefont {Z.}~\bibnamefont {Niu}}, \bibinfo
  {author} {\bibfnamefont {S.}~\bibnamefont {Hughes}}, \bibinfo {author}
  {\bibfnamefont {X.}~\bibnamefont {Wang}},\ and\ \bibinfo {author}
  {\bibfnamefont {J.}~\bibnamefont {Liu}},\ }\bibfield  {title} {\bibinfo
  {title} {Dynamic resonance fluorescence in solid-state cavity quantum
  electrodynamics},\ }\bibfield  {journal} {\bibinfo  {journal} {Nature
  Photonics}\ }\href {https://doi.org/10.1038/s41566-023-01359-x}
  {10.1038/s41566-023-01359-x} (\bibinfo {year} {2024})\BibitemShut {NoStop}%
\bibitem [{\citenamefont {Boos}\ \emph {et~al.}(2024)\citenamefont {Boos},
  \citenamefont {Kim}, \citenamefont {Bracht}, \citenamefont {Sbresny},
  \citenamefont {Kaspari}, \citenamefont {Cygorek}, \citenamefont {Riedl},
  \citenamefont {Bopp}, \citenamefont {Rauhaus}, \citenamefont {Calcagno},
  \citenamefont {Finley}, \citenamefont {Reiter},\ and\ \citenamefont
  {M\"uller}}]{PhysRevLett.132.053602}%
  \BibitemOpen
  \bibfield  {author} {\bibinfo {author} {\bibfnamefont {K.}~\bibnamefont
  {Boos}}, \bibinfo {author} {\bibfnamefont {S.~K.}\ \bibnamefont {Kim}},
  \bibinfo {author} {\bibfnamefont {T.}~\bibnamefont {Bracht}}, \bibinfo
  {author} {\bibfnamefont {F.}~\bibnamefont {Sbresny}}, \bibinfo {author}
  {\bibfnamefont {J.~M.}\ \bibnamefont {Kaspari}}, \bibinfo {author}
  {\bibfnamefont {M.}~\bibnamefont {Cygorek}}, \bibinfo {author} {\bibfnamefont
  {H.}~\bibnamefont {Riedl}}, \bibinfo {author} {\bibfnamefont {F.~W.}\
  \bibnamefont {Bopp}}, \bibinfo {author} {\bibfnamefont {W.}~\bibnamefont
  {Rauhaus}}, \bibinfo {author} {\bibfnamefont {C.}~\bibnamefont {Calcagno}},
  \bibinfo {author} {\bibfnamefont {J.~J.}\ \bibnamefont {Finley}}, \bibinfo
  {author} {\bibfnamefont {D.~E.}\ \bibnamefont {Reiter}},\ and\ \bibinfo
  {author} {\bibfnamefont {K.}~\bibnamefont {M\"uller}},\ }\bibfield  {title}
  {\bibinfo {title} {Signatures of dynamically dressed states},\ }\href
  {https://doi.org/10.1103/PhysRevLett.132.053602} {\bibfield  {journal}
  {\bibinfo  {journal} {Phys. Rev. Lett.}\ }\textbf {\bibinfo {volume} {132}},\
  \bibinfo {pages} {053602} (\bibinfo {year} {2024})}\BibitemShut {NoStop}%
\bibitem [{\citenamefont {Arranz~Regidor}\ \emph {et~al.}(2024)\citenamefont
  {Arranz~Regidor}, \citenamefont {Knorr},\ and\ \citenamefont
  {Hughes}}]{sofia2024}%
  \BibitemOpen
  \bibfield  {author} {\bibinfo {author} {\bibfnamefont {S.}~\bibnamefont
  {Arranz~Regidor}}, \bibinfo {author} {\bibfnamefont {A.}~\bibnamefont
  {Knorr}},\ and\ \bibinfo {author} {\bibfnamefont {S.}~\bibnamefont
  {Hughes}},\ }\bibfield  {title} {\bibinfo {title} {Theory and simulations of
  few-photon fock state pulses strongly interacting with a single qubit in a
  waveguide: exact population dynamics and time-dependent spectra}} (\bibinfo
  {year} {2024}),\ \bibinfo {note} {manuscript in preparation}\BibitemShut
  {NoStop}%
\end{thebibliography}%

\end{document}